%% file: tmi_sir_ano.tex
\documentclass[journal]{IEEEtran}
\usepackage{graphicx}
\usepackage{amsmath}
\usepackage{amssymb}
\usepackage{color}
\usepackage[letterpaper=true,colorlinks,bookmarks=false]{hyperref}
\usepackage{subfigure}
\usepackage[lined,boxed, ruled]{algorithm2e}
\usepackage{url}

\usepackage{hyperref}

\usepackage{multirow} % table
\usepackage[table,xcdraw]{xcolor} % table

\input{shared_defs}

\def\orcidzhoukang{\href{https://orcid.org/0000-0001-8789-4243}{\includegraphics[scale=0.06]{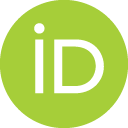}}}
\def\orcidlijing{\href{https://orcid.org/0000-0001-7027-7513}{\includegraphics[scale=0.06]{figures/ORCIDiD.png}}}
\def\orcidluowx{\href{https://orcid.org/0000-0002-0754-6458}{\includegraphics[scale=0.06]{figures/ORCIDiD.png}}}
\def\orcidyangjl{\href{https://orcid.org/0000-0001-7450-1260}{\includegraphics[scale=0.06]{figures/ORCIDiD.png}}}
\def\orcidfuhz{\href{https://orcid.org/0000-0002-9702-5524}{\includegraphics[scale=0.06]{figures/ORCIDiD.png}}}
\def\orcidchengjun{\href{https://orcid.org/0000-0003-1786-6188}{\includegraphics[scale=0.06]{figures/ORCIDiD.png}}}
\def\orcidliujiang{\href{https://orcid.org/0000-0001-6281-6505}{\includegraphics[scale=0.06]{figures/ORCIDiD.png}}}
\def\orcidgaoshh{\href{https://orcid.org/0000-0003-1626-2040}{\includegraphics[scale=0.06]{figures/ORCIDiD.png}}}

\begin{document}

%\title{SIR-Ano: Superpixel-image-bridged Image Reconstruction Network for Anomaly Detection}
%\title{Reconstruction with Intermediate Proxy \\ for Anomaly Detection in Medical Image}
\title{\highre{Proxy-bridged} Image Reconstruction Network for Anomaly Detection \highre{in Medical Images}}

%\author{Authors
\author{
	Kang Zhou\orcidzhoukang, Jing Li\orcidlijing, Weixin Luo\orcidluowx, Zhengxin Li, Jianlong Yang\orcidyangjl, \\ Huazhu Fu\orcidfuhz, Jun Cheng\orcidchengjun, Jiang Liu\orcidliujiang and Shenghua Gao\orcidgaoshh
\thanks{Corresponding Author: Shenghua Gao, gaoshh@shanghaitech.edu.cn.}
\thanks{Kang Zhou, Jing Li, Weixin Luo, Zhengxin Li and Shenghua Gao are with School of Information Science and Technology, ShanghaiTech University, Shanghai 201210, China; Kang Zhou and Jing Li are also with Shanghai Institute of Microsystem and Information Technology, Chinese Academy of Sciences, Shanghai 200050, China, and also University of Chinese Academy of Sciences, China. Kang Zhou and Jing Li contributed equally to this work. Email: \{zhoukang, lijing1\}@shanghaitech.edu.cn.
Shenghua Gao is also with Shanghai Engineering Research Center of Intelligent Vision and Imaging, Shanghai 201210, China, and also with Shanghai Engineering Research Center of Energy Efficient and Custom AI IC, Shanghai 201210, China.
}
\thanks{
	Jianlong Yang is with School of Biomedical Engineering, Shanghai Jiao Tong University, Shanghai 200030, China.}
\thanks{
	Huazhu Fu is with the Institute of High Performance Computing (IHPC), Agency for Science, Technology and Research (A*STAR), Singapore 138632.
%	Inception Institute of Artificial Intelligence, Abu Dhabi, United Arab Emirates.
}
\thanks{
	Jun Cheng is with the Institute for Infocomm Research, Agency for Science, Technology and Research (A*STAR), Singapore 138632.}
\thanks{
	Jiang Liu is with the Department of Computer Science and Engineering, Southern University of Science and Technology, Guangdong 518055, China, and also with Cixi Institute of Biomedical Engineering, Chinese Academy of Sciences, Zhejiang 315201, China. }
}

%\markboth{}%
%{Zhou \MakeLowercase{\textit{et al.}}
%}

\maketitle
\begin{abstract}
Anomaly detection in medical images refers to the identification of abnormal images with only normal images in the training set.
Most existing methods solve this problem with a self-reconstruction framework, which tends to learn an identity mapping and reduces the sensitivity to anomalies. To mitigate this problem, in this paper, we propose a novel \highre{Proxy-bridged Image Reconstruction Network (ProxyAno)} for anomaly detection in medical images. \highre{Specifically, we use an intermediate proxy to bridge the input image and the reconstructed image. We study different proxy types, and we find that the superpixel-image (SI) is the best one. We set all pixels' intensities within each superpixel as their average intensity, and denote this image as SI.} The proposed ProxyAno consists of two modules, a Proxy Extraction Module and an Image Reconstruction Module. In the Proxy Extraction Module, a memory is introduced to memorize the feature correspondence for normal image to its corresponding SI, while the memorized correspondence does not apply to the abnormal images, which leads to the information loss for abnormal image and facilitates the anomaly detection. In the Image Reconstruction Module, we map an SI to its reconstructed image. Further, we crop a patch from the image and paste it on the normal SI to mimic the anomalies, and enforce the network to reconstruct the normal image even with the pseudo abnormal SI. In this way, our network enlarges the reconstruction error for anomalies. Extensive experiments on brain MR images, retinal OCT images and retinal fundus images verify the effectiveness of our method for both image-level and pixel-level anomaly detection. 
%Further, as far as we know, this is the first work that verifies the effectiveness of anomaly detection for abnormal region segmentation for retinal fundus images.
\end{abstract}


\IEEEpeerreviewmaketitle

\input{sections/intro}
\input{sections/related}

\input{sections/method}

\input{sections/exps}
%\input{sections/discussion}

\section{Conclusion}
\label{sec_clu}
In this paper, we propose a novel ProxyAno approach to mitigate the identity mapping issue in the Auto-Encoder based anomaly detection paradigm. Specifically, our ProxyAno first maps an input image to a superpixel-images based on a memory-aided Proxy Extraction Module, and this memory would cause information loss for abnormal images and facilitate the anomaly detection. Further, an Image Reconstruction Module is used to reconstruct the input based on the superpixel-images. To further enlarge the reconstruction error, we enforce the reconstruction module can well reconstruct the input even with a pseudo abnormal superpixel-image. In this way, our ProxyAno favors the reconstruction of normal input and leads to a large reconstruction of the abnormal images. Extensive experiments on different medical image datasets validate the effectiveness of our approach for both image-level and pixel-level anomaly detection.

\section{Acknowledge}
The work was supported by National Key R\&D Program of China (2018AAA0100704), NSFC \#61932020,  Science and Technology Commission of Shanghai Municipality (Grant No. 20ZR1436000), 
Guangdong Provincial Department of Education (2020ZDZX3043), Shenzhen Natural Science Fund (JCYJ20200109140820699), the Stable Support Plan Program (20200925174052004), and ``Shuguang Program" supported by Shanghai Education Development Foundation and Shanghai Municipal Education Commission. 
We would like to thank the Editor-in-Chief, associate editor, and all anonymous reviewers for their valuable and constructive comments.

%\vspace{-0.1in}

\bibliographystyle{IEEEtran}
\bibliography{/latex_shared_files/zhou_anomaly_detection}

\end{document}

%% file: shared_defs.tex
\usepackage{amssymb}

\newcommand{\highre}[1]{{\textcolor{black}{#1}}} % Clean Copy
\newcommand{\HILIGHT}[1]{{\textcolor{black}{#1}}} % Minor Revision

\newcommand{\first}[1]{{\textbf{\textcolor{red}{\underline{#1}}}}}
\newcommand{\second}[1]{{\textcolor{red}{#1}}}
\def\etal{\emph{et al. }}
\def\ie{\emph{i.e.}}
\def\eg{\emph{e.g.}}
\def\cL{{\mathcal L}}

\def\bR{{\mathbb R}}

\def\bI{{\mathbf I}}
\def\bhI{{\hat{\mathbf I}}}
\def\bhIb{{\hat{\mathbf I ^{'}}}}
\def\bP{{\mathbf P}}
\def\bPb{{\mathbf P ^{'}}}

\def\ben{{\textbf{\textit{Enc}}}}
\def\bde{{\textbf{\textit{Dec}}}}
\def\cf{{\mathcal{F}}}
\def\cg{{\mathcal{G}}}

\def\cL{{\mathcal L}}

\def\bE{{\mathbb E}}
\def\bD{{\mathbf D}}

\def\bM{{\mathbf M}}
\def\bA{{\mathbf A}}

%% file: sections/intro.tex
\section{Introduction}
Anomaly detection aims to identify abnormalities with only normal data in the training set \cite{chandola2009anomaly,pang2021deep,zhou2021memorizing}. Recently it has drawn much attention in the community of medical image analysis \cite{schlegl2017unsupervised, zimmerer2019unsupervised, baur2020bayesian, chen2020unsupervised}. 
There are three reasons for this: firstly, it is not easy to acquire medical data, especially for some rare diseases; secondly, it is expensive to annotate the lesion; thirdly, it is comparatively easier to collect the data from healthy subjects. 
Because of the data constraint, anomaly detection in medical images is more challenging than disease classification that has been well tackled with deep convolutional neural networks (CNNs) \cite{li2019canet, luo2020deep, srinivasan2014fully, zhou2018multi}.

To tackle the anomaly detection with only normal data, \highre{many methods have been proposed. These methods can be roughly categorized into two groups. The first group is reconstruction-based method, which train a model to reconstruct the normal image, and in the test phase, the reconstruction error of abnormal images is larger than the normal ones.
The second group is non-reconstruction-based methods. For example, Ouardini \etal \cite{ouardini2019towards} proposed an efficient and effective transfer-learning based approach for anomaly detection on retinal fundus images. Golan \etal \cite{golan2018deep} proposed to learn a meaningful representation of the learned training data in a fully discriminative fashion using the self-labeled dataset.}

\highre{
In this work, we focus on the study of reconstruction-based anomaly detection. Most previous reconstruction-based methods follow a self-reconstruction paradigm.
}
Specifically, an Auto-Encoder (AE) is commonly used for anomaly detection \cite{baur2018deep, zhou2017anomaly}. In the training phase, the AE is learnt with only normal data, and it is expected that the reconstructed image from the decoder will be close to the input for the normal image, and the output is different from the input for the abnormal images in the test phase. 
In addition, generative adversarial network (GAN) \cite{goodfellow2014generative} based approaches have also be introduced to guarantee the fidelity of the reconstruction for normal data \cite{chen2018unsupervised, schlegl2017unsupervised}. Further, other than measuring the anomaly with reconstruction error in the image space, Akcay \etal \cite{akcay2018ganomaly} proposed to append another encoder to extract latent features corresponding to the reconstructed image, and take the difference of latent features corresponding to the input and reconstructed image as the anomaly measurement. 
\highre{However, it is worth noting that self-reconstruction is essentially to learn an identity mapping function due to the information equivalence, \ie, the equivalence between the input and output of the model during the training \cite{fei2020attribute}. Specifically, self-reconstruction aims to reproduce the output to approximate the input. They may hence tend to overfit to learn an identity mapping between the input and output \cite{steck2020autoencoders}. Therefore, self-reconstruction-based anomaly detection methods cannot guarantee the large reconstruction error for abnormal images, such that it may not be sensitive to anomalies.}

\begin{figure*}[ttt]
	\centering
	\includegraphics[width=1\textwidth]{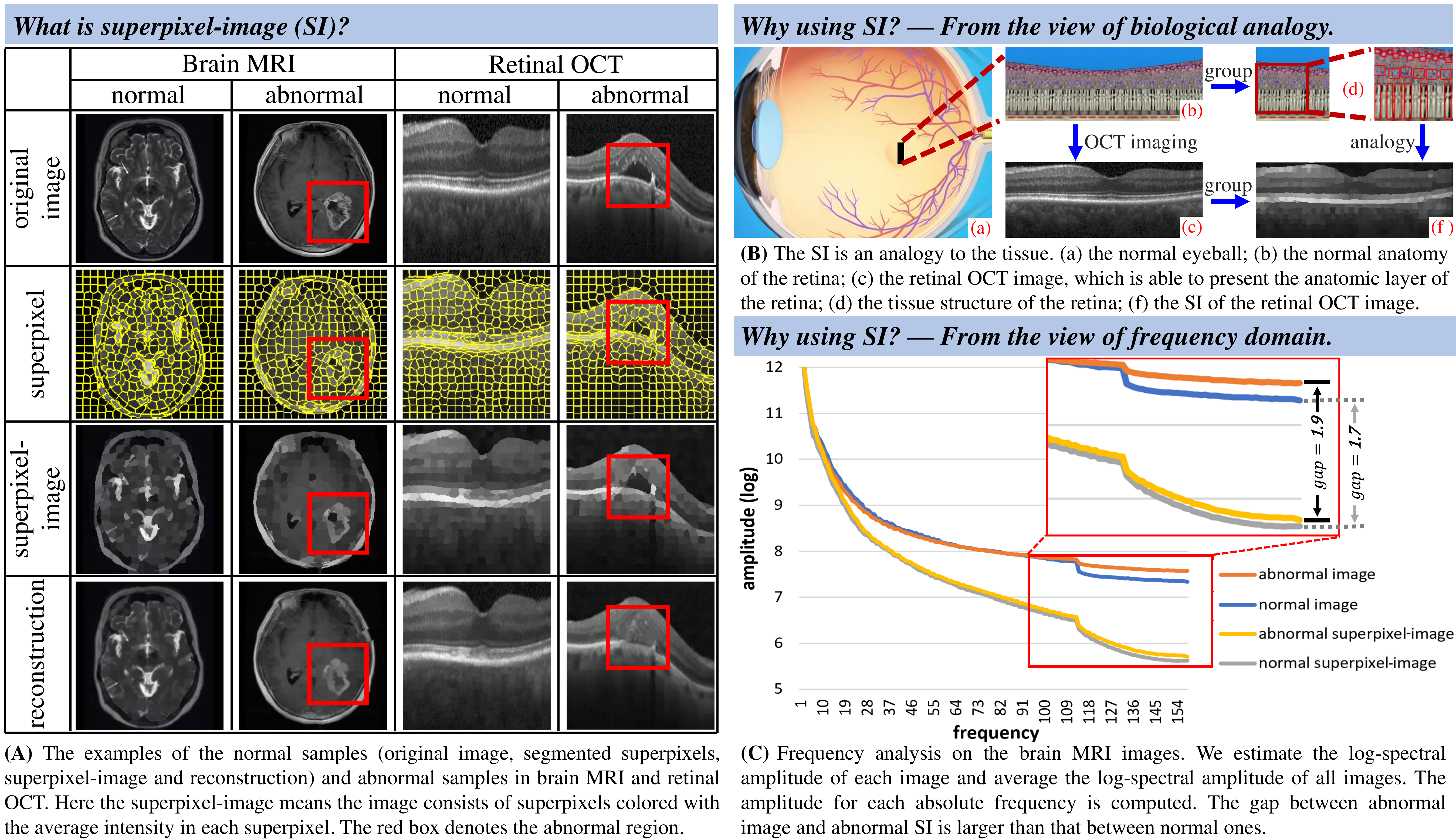}
	\vspace{-0.2in}
	\caption{
	\highre{The definition of SI and the motivations of using SI as the intermediate proxy to bridge the input image and the reconstructed image.}
}
	\vspace{-0.15in}
	\label{fig1}
\end{figure*}

%\begin{figure}[ttt]
%	\centering
%	\includegraphics[width=.48\textwidth]{figures/fig_intro}
%	\vspace{-0.1in}
%	\caption{
%		The examples of the normal samples (original image, segmented superpixels, superpixel-image and reconstruction) and abnormal samples in brain MRI and retinal OCT.
%		Here the superpixel-image (SI) means the image consists of superpixels colored with the average intensity in each superpixel.
%	}
%	\vspace{-0.1in}
%	\label{fig1}
%\end{figure}

To mitigate the identity mapping problem in self-reconstruction-based anomaly detection, \highre{we use an intermediate proxy to bridge the input image and the reconstructed image.
We study different proxy types (\eg, edge, smooth image), and we find that the superpixel-image (SI) is the best one. 
As shown in Fig. \ref{fig1} (A), we set all pixels' intensities within each superpixel to their average intensity for each superpixel and denote this image as SI.
In our proposed anomaly detection method, we map the input image to a proxy with the Proxy Extraction Module, and map the proxy to its reconstructed image with the Image Reconstruction Module. 
We name our method as Proxy-bridged Image Reconstruction Network (ProxyAno).}

%\todo{why using SI?}
%\highre{
\textbf{The motivations behind using SI as the proxy are the following aspects.} \textbf{1) From the view of biomedical analogy.} 
First of all, we would like to argue that the SI is an analogy to the tissue.
Concretely, the scale of tissue is between that of the cell and the organ, and the tissue is a group of many similar cells and carries out a specific function. Similarly, the superpixel is defined as a group of pixels with similar characteristics (\eg, intensity) and the scale of superpixel is between that of the pixel and the image \cite{ren2003learning}. 
As shown in Fig. \ref{fig1} (B), we take the eyeball as the specific example to illustrate  the analogy.
When a disease occurs, the cells and tissues in the lesion region undergo pathological changes \cite{kinane2001causation, nosalski2017perivascular, gutmann1996acute}. In other words, the tissue-level pathological change represents the disease occurring.
Analogously, the SI-level change partly represents the anomaly that occurs in the given image \cite{li2020superpixel}.
This analogy inspires us to use the SI as an intermediate proxy for image reconstruction.
\textbf{2) From the view of frequency domain.}
In medical images, the average gradient (\ie, frequency) of the abnormal region is \HILIGHT{usually} larger than that of the normal region \cite{zimmerer2019context, baur2021autoencoders}. 
\HILIGHT{By transforming the original image to its SI, the abnormal region loses more information than the normal region. Thus, the variation degree of the abnormal region is more significant than that of the normal region.
Therefore, by leveraging SI as the intermediate proxy, the reconstruction from SI of the abnormal samples is more difficult than that of the normal samples, leading to favoring detecting anomalies.
We take a specific example in Fig. \ref{fig1} (C). 
The images in the brain MRI dataset \cite{rai2019hybrid} are transferred 
with 2D Fourier Transform \cite{bracewell2004fourier} and the log-spectral amplitude is estimated. 
It can be observed that the major distinction between normal images and abnormal images is the high-frequency information,
and the decrease degree of the high-frequency's amplitude of the abnormal samples is more significant than that of the normal samples.}
\textbf{3) From the definition of superpixel.} The definition of superpixel determines that the SI can preserve both the local structure and texture information \cite{forsyth2002computer, mori2004recovering}. In other words, SI contains mid-level semantic information \cite{kwak2017weakly}.
%}

\highre{
By using the SI as the intermediate proxy to bridge the input image and the reconstructed image, it is expected that the network will lead to a large prediction error in SI for an abnormal image, and consequently will lead to a large reconstruction error for an abnormal image.}
%, which would favor the anomaly detection. 
%In contrast, the reconstruction error is relatively small for normal images. 
%In this way, the proposed SIR-Ano mitigates the identity mapping issue and preserve the sensitivity to abnormal images. Additionally, our 
The proposed ProxyAno consists of two modules, \ie, a Proxy Extraction Module to map the input image to an SI, and an Image Reconstruction Module to map the SI to the reconstructed image.
Specifically, in the Proxy Extraction Module, we introduce a memory to memorize the mapping pattern from the input image to its corresponding SI for normal training data. Given an input, we use its latent feature extracted by the encoder in the Proxy Extraction Module to retrieve the most similar item in the memory, and feed the retrieved feature into the decoder in Proxy Extraction Module to predict the SI. 
Then we feed the predicted SI into the Image Reconstruction Module to reconstruct the input image. 
To further enlarge the reconstruction error for abnormal images, we create pseudo abnormal SI by cutting a patch from a randomly selected normal image and pasting the patch on the normal proxy. Other than the reconstruction from the normal SI, we also enforce the network well reconstruct the input even with the pseudo abnormal SI. 
In this way, the Image Reconstruction Module would have the ability to repair the anomaly, which would result in a large reconstruction error for abnormal images. Experiments on the images of different modalities validate the effectiveness of our approach for both image-level and pixel-level anomaly detection.

The main contributions are summarized as follows:

\begin{enumerate}
	\item To mitigate the identity mapping problem in self-reconstruction-based anomaly detection, under the assumption that the mapping from an image and its SI can be predictable for the normal images while the mapping for abnormal images cannot be predicted, we propose the ProxyAno to connect the input image and the reconstructed image with an SI.
	\item In the Proxy Extraction Module, we introduce a memory to memorize the correspondence between the input image and its SI for normal images, while the memorized correspondence does not apply to the abnormal images. 
	\item In the Image Reconstruction Module, a pseudo abnormal SI is built to enforce the network to reconstruct the normal input even with the abnormal SI. This would further enlarge the reconstruction error for abnormal image.
	\item Extensive experiments on three modalities (\ie, brain MRI, retinal OCT and retinal fundus) demonstrate the effectiveness of our approach for anomaly detection. 
	Moreover, \highre{our work verifies the effectiveness} of anomaly detection for abnormal region segmentation task in retinal fundus images.
\end{enumerate}

The rest of this paper is organized as follows:
In Section \ref{sec_related}, we introduce the work related to anomaly detection in medical images.
In Section \ref{sec_method}, we describe our proposed ProxyAno for anomaly detection in detail.
In Section \ref{sec_exp}, extensive experiments on three modalities are conducted to validate the effectiveness of our method.
We conclude our work in Section \ref{sec_clu}.

%% file: sections/related.tex
\section{Related Work}
\label{sec_related}

%\subsection{Anomaly Detection in Medical Images}
Recently, anomaly detection has drawn much attention in the medical image domain, including different imaging modalities, such as brain MRI \cite{zimmerer2019unsupervised, baur2020bayesian, chen2020unsupervised, baur2018deep, zimmerer2018context,  han2020madgan}, retinal OCT \cite{schlegl2017unsupervised, schlegl2019f, zhou2020sparse, seebock2019exploiting}, chest X-Ray \cite{zhang2020covid, tang2019abnormal, bozorgtabar2020salad, tuluptceva2020anomaly} and brain CT \cite{pawlowski2018unsupervised}.
\highre{In this paper, we focus on the reconstruction-based anomaly detection in medical images.}
These works can be roughly categorized into two categories: image-level anomaly detection and pixel-level anomaly detection.
\highre{
The pixel-level anomaly detection also called as abnormal region segmentation and anomaly segmentation, which classifies the pixels in the given image into normal or abnormal. 
}

\begin{figure*}[ttt]
	\centering
	\includegraphics[width=1\textwidth]{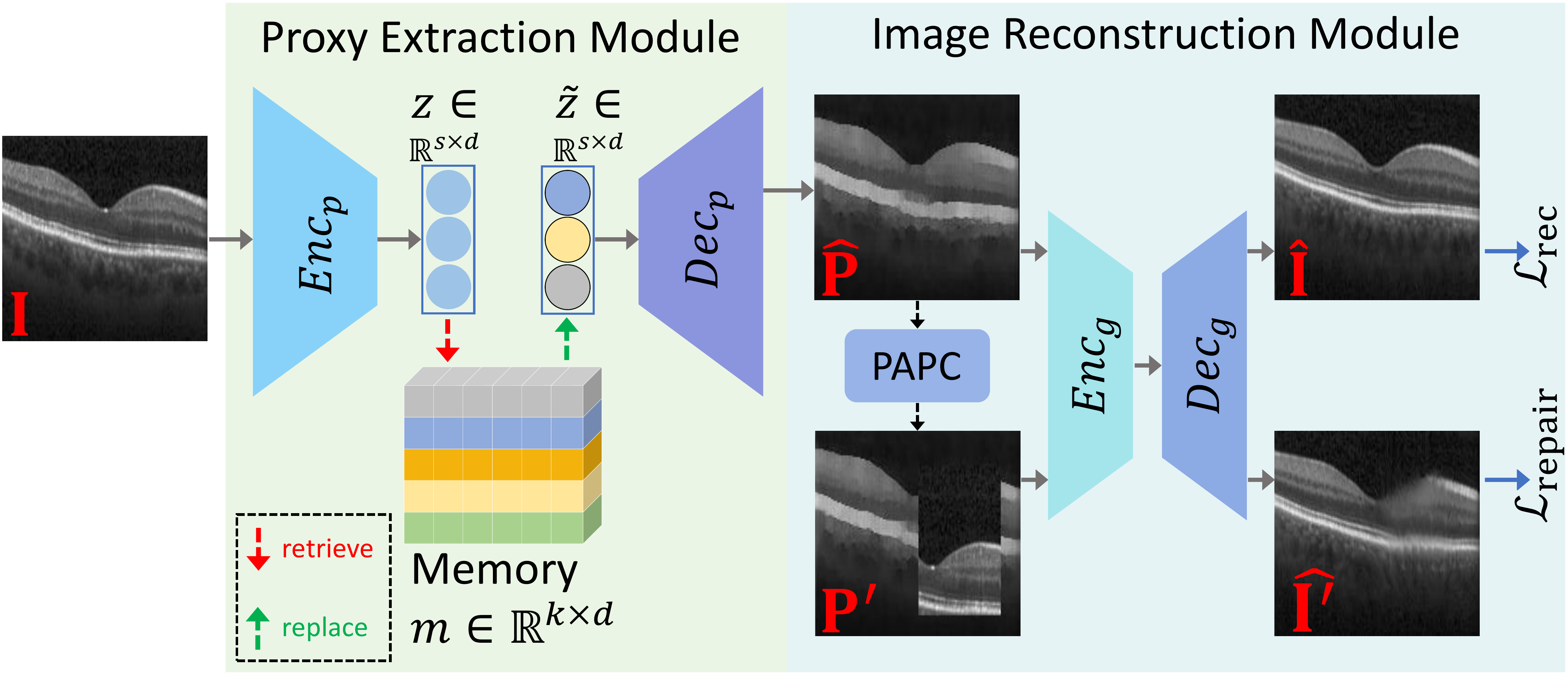}
	\vspace{-0.15in}
	\caption{
		The overview of the proposed network, which consists of two modules, a Proxy Extraction Module and an Image Reconstruction Module. 
		PAPC is the short for the Pseudo Abnormal Proxy Constructor, and PAPC only works in the training phase. $\hat{\bP}$ and $\bPb$ denote the predicted proxy and pseudo abnormal proxy, respectively.
	}
	\label{fig_overall}
	\vspace{-0.1in}
\end{figure*}

\textbf{Image-Level Anomaly Detection.}
Most previous image-level anomaly detection methods are based on a self-reconstruction framework that assumes the abnormal images cannot be well reconstructed by a model learned merely with the normal images.
Specifically, Zimmerer \etal proposed to use a variational auto-encoder for the brain MRI image reconstruction \cite{zimmerer2019unsupervised}, and they also proposed to combine a context-encoder and a variational auto-encoder for the brain MRI image reconstruction \cite{zimmerer2018context}.
Schlegl \etal \cite{schlegl2017unsupervised} proposed to utilized a GAN \cite{goodfellow2014generative} for image reconstruction. They first trained a generator mapping the latent vector to the reconstructed image. Then they fixed the generator and the discriminator to train a mapping from the input image to a latent vector with the back-propagation algorithm.
Based on \cite{schlegl2017unsupervised}, Han \etal \cite{han2020madgan} proposed to take multiple adjacent slices of 3D MRI data as the input for the reconstruction. 
Further, to address the slow mapping issue from an image to a latent vector in \cite{schlegl2017unsupervised}, Schlegl \etal \cite{schlegl2019f} proposed to use an encoder to map the input image to the latent vector.
Inspired by the success of pix2pix \cite{isola2017image}, which trains an encoder and a generator with an end-to-end manner, Zhou \etal \cite{zhou2020sparse} proposed to use the pix2pix and regularize the sparsity of the latent space (termed Sparse-GAN) to guarantee the fidelity of the reconstruction for normal data.
\highre{Ouardini \etal \cite{ouardini2019towards} proposed to employ the Inception-ResNet-v2 network \cite{szegedy2017inception} trained on ImageNet dataset \cite{russakovsky2015imagenet} as a feature extractor, and subsequently feeds the extracted feature into Isolation Forests \cite{liu2008isolation}, which is a robust and efficient anomaly detection method. They applied their method to image-level anomaly detection on retinal fundus image and conducted experiments on two datasets, which contain Retinopathy of Prematurity (ROP) and Diabetic Retinopathy (DR), respectively. The model \cite{ouardini2019towards} can classify the given image into normal or abnormal (ROP and DR).}
Very recently, to leverage the structure information (\eg, the retinal blood vessels in fundus image, and the retinal layers in OCT image) for anomaly detection, Zhou \etal \cite{zhou2020encoding} proposed a P-Net to reconstruct the image, which leverages the relation between structure and texture for image-level anomaly detection. 
\highre{
Zhou \etal \cite{zhou2020encoding} apply image-level anomaly detection to multiple retinal disease detection. Since image-level anomaly detection is a two-classes classification problem, they regard the images with DR, glaucoma, age-related macular degeneration and pathological myopia as the abnormal images. Then the P-Net \cite{zhou2020encoding}  classifies the given image into normal or abnormal.
}

\textbf{Pixel-Level Anomaly Detection.}
Besides used for image-level anomaly detection, reconstruction based methods are also commonly used for pixel-level anomaly detection. Most existing methods train a reconstruction model with the normal data, and in the test phase, the residual map (a.k.a anomaly map, lesion map) is obtained by subtract reconstructed image with the input image.
Chen \etal \cite{chen2018unsupervised} initially proposed to use adversarial auto-encoders for anomaly segmentation on brain MRI images. Almost the same time, Baur \etal \cite{baur2018deep} proposed to use a deep auto-encoder that combines spatial AEs and GANs for anomaly segmentation on brain MRI.
In addition, the Bayesian deep learning also been introduced for the pixel-level anomaly detection.
Pawlowski \etal \cite{pawlowski2018unsupervised} proposed to use the Bayesian auto-encoder to model the normal data distribution and apply the algorithm for anomaly detection on brain CT.
%Seebock \etal \cite{seebock2019exploiting} introduced a Bayesian U-Net that exploits segmentation models of normal anatomy and their epistemic uncertainty on the test images. 
Seebock \etal \cite{seebock2019exploiting} introduced a Bayesian U-Net that exploits the segmentation model of normal anatomy and its epistemic uncertainty for anomaly segmentation on the test images. 
Very recently, Chen \etal \cite{chen2020unsupervised} proposed a probabilistic model that uses a network-based prior as the normative distribution and detects lesions using Maximum-A-Posteriori estimation.
Baur \etal \cite{baur2020bayesian} further proposed a dropout-U-Net based auto-encoder, which introduces the dropout connections between the encoder and decoder, to model the uncertainty and enable high fidelity reconstructions on brain MRI images. 

\textbf{Summary.} In a summary, almost all previous works adopted the self-reconstruction paradigm for anomaly detection \highre{in medical images}.
As aforementioned, the self-reconstruction model is essentially to learn an identity mapping function, which cannot guarantee a large reconstruction error for abnormal images.
To mitigate the identity mapping problem, we propose to introduce an SI as the proxy to bridge the input image and the reconstructed image. 
\highre{Moreover, different with \cite{zhou2020encoding, ouardini2019towards}, which conducted image-level anomaly detection (\ie, classifying an image into normal or abnormal) on the retinal fundus image, this work verifies the effectiveness of pixel-level anomaly detection on retinal fundus images. To the best of our knowledge, we are the first to conduct anomaly segmentation in the retinal fundus modality.}

%% file: sections/method.tex
\section{Method}
\label{sec_method}
\subsection{Overview}
In the reconstruction based anomaly detection framework, it is expected that the network can well reconstruct the normal input image, and the reconstructed image of the abnormal ones would be with a large reconstruction error. In this way, the abnormal images can be distinguished from the normal images. 
Towards this end, in this paper, we introduce a Proxy-bridged Image Reconstruction Network (ProxyAno) for anomaly detection in medical images. The proposed ProxyAno consists of two modules, a Proxy Extraction Module $\cf_p(\cdot)$ and an Image Reconstruction Module $\cf_g(\cdot)$. Specifically, the proposed ProxyAno can be formulated as:
\begin{equation}
%\vspace{-0.2in}
\begin{aligned}
\hat{\bP} &= \cf_p(\bI) \\
\bhI &= \cf_g(\hat{\bP})
\end{aligned}
%\vspace{-0.1in}
\end{equation}
where $\hat{\bP}$ denotes the predicted SI, $\bI$ and $\bhI$ denote original image and reconstructed image, respectively.
We use $\ben_p$/$\bde_p$ to denote the encoder/decoder in the Proxy Extraction Module, and use $\ben_g$/$\bde_g$ to denote the encoder/decoder in the Image Reconstruction Module, respectively.

Fig. \ref{fig_overall} illustrates the overview of our ProxyAno. To enlarge the reconstruction error for the abnormal image while maintaining a small reconstruction error for the normal images, we propose to:
1)
memorize the mapping pattern between the normal input image and its corresponding SI with a memory in the Proxy Extraction Module. Once this module is well trained, the weights of this module are fixed when training the Image Reconstruction Module;
2)
reconstruct the image from its SI, meanwhile repairing the abnormal SI in the Image Reconstruction Module.
To achieve this, we first create a pseudo abnormal proxy $\bPb$ and feed it into the Image Reconstruction Module with $\bhIb = \cf_g(\bPb)$.
Then we propose a repairing loss to enforce the similarity between the reconstructed image $\bhIb$ of the pseudo abnormal proxy and the normal image $\bI$.
Benefiting from these two improvements, in the test phase, both $\hat{\bP}$ and $\bhI$ tend to be close to the normal ones, leading to a small reconstruction error for the normal samples and a large reconstruction error on the abnormal ones.
Then the reconstruction error is used as a measurement to detect the anomalies.

\subsection{Superpixel-image-bridged Image Reconstruction Network}

%\begin{figure}[htb]
%	\centering
%	\includegraphics[width=.49\textwidth]{figures/fig_superpixel}
%	\vspace{-0.25in}
%	\caption{
%			The SI is an analogy to the tissue in the retinal OCT image.
%		(a) the normal eyeball; (b) the normal anatomy of the retina; (c) the retinal OCT image, which is able to present the anatomic layer of the retina; (d) the tissue structure of sub-image (b); (f) the SI of sub-image (c), and SI is used as the proxy in our method.
%	}
%	\label{fig_superpixel}
%\end{figure}

In our work, we use the SLIC \cite{achanta2012slic} algorithm to obtain the superpixels.
The SLIC algorithm groups meaningful pixels into a superpixel by composing spatially adjacent pixels.
Based on the superpixels, we can get an SI that consists of the superpixel colored with the average intensity for pixels within each superpixel. 
Then we use the SI extracted with the SLIC algorithm as supervision to train a Proxy Extraction Module. 
It is worth noting that directly applying the SLIC algorithm on the abnormal image, the superpixel structure of lesion can also be extracted. Thus, the abnormal SI contains some lesion information; consequently, the lesion in the abnormal images may also be reconstructed. To increase the fidelity of the reconstruction of the normal images and increase the reconstruction error for the abnormal ones, on the one hand, we introduce a memory to map the abnormal image to a normal SI, on the other hand, we also propose to let the algorithm automatically repairing an SI of an abnormal image, and map it to a normal input image.

\subsection{Proxy Extraction Module}
\label{memory}

In the Proxy Extraction Module, we propose to \highre{augment the network with a memory to explicitly memorize} the correspondence between the normal input and its SI. 
\highre{The motivations behind using a memory to augment the network are three aspects: 1) to recognize disease in medical images, the clinicians need to memorize the characteristics of normal samples, and the superpixel structure is an important characteristic in the image;
2) the memory is a type of location-independent long-term attention, which augments the ability of convolutional structure;
3) the correspondence between the normal input and its SI is very regular. Using the correspondence patterns is probably enough to generate all normal SI's from its input.
Therefore,} we introduce a memory that can memorize the normal patterns. 
\highre{The feature is first extracted with a encoder. Then, instead directly feed the feature into the decoder, we use the feature as a query to retrieve the most relevant item in the memory. The features feed into the decoder are obtained from a selected memory item of the normal data.}

Since this memory is learnt based on the correspondence of normal images, it can be expected that this would cause information loss for the abnormal ones when generating its SI. Consequently, the image reconstruction form this SI would be with a large reconstruction error, which is a desirable property for anomaly detection. 
To achieve this goal, we introduce a memory $m \in \bR ^ {k \times d}$ to memorize the latent features in the Proxy Extraction Module, where $k$ is the size (the number of items) of the memory , and $d$ is the dimensionality of the latent vector $m_j \in \bR ^ d, j \in \{1,2, \cdots, k\}$.
As shown in Fig. \ref{fig_overall}, the framework takes an image $\bI$ as an input, and then $\bI$ is passed through the encoder $\ben_p$ to extract a latent feature $z\in \bR ^ {h \times w \times d}$, where $h$ and $w$ are the height and the width of the feature map. In this module, $z$ is used to search for the nearest neighbor item in the memory $m$, and we denote the nearest item of $z$ as $\tilde{z} \in \bR ^ {s \times d}$, here $\tilde{z}$ indicates the nearest item in the memory of a feature. 

\begin{figure}[htb]
	\vspace{-0.1in}
	\centering
	\includegraphics[width=.5\textwidth]{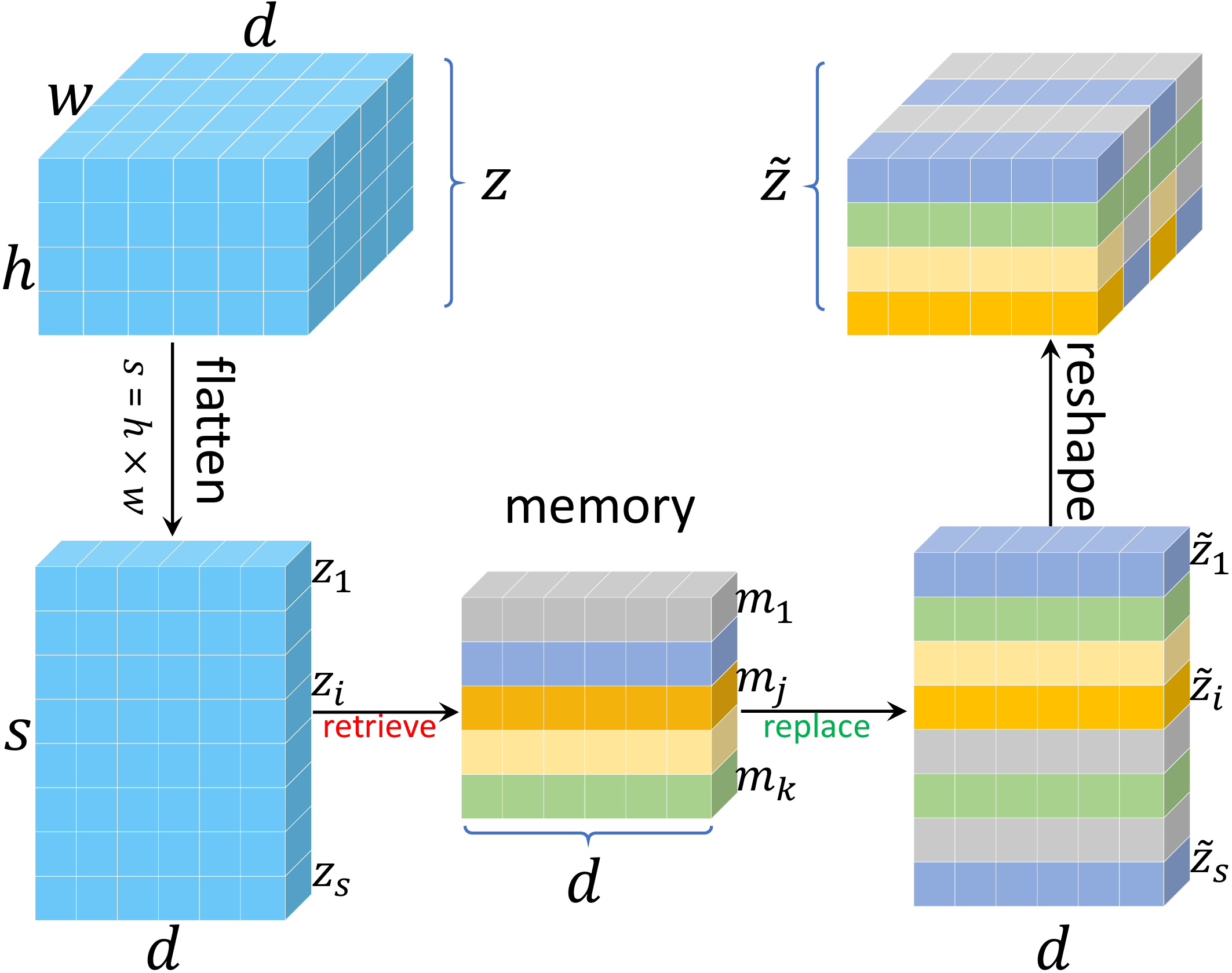}
	\vspace{-0.1in}
	\caption{The illustration of the detailed operation in memory. $z_i \in \bR ^ d$ retrieves the nearest vector $m_\text{j} \in \bR ^ d$ in the memory, and $z_i$ is replaced by $\tilde{z}_i$ with Equation (\ref{equ_memory_replace}).
	}
	\label{fig_memory}
\end{figure}

Specifically, as illustrated in Fig. \ref{fig_memory},
we first flatten the latent feature as $z\in \bR ^ {s \times d}$, where $s = h \times w $.
Then $\forall  i \in \{1, \cdots, s\}$, $z_i \in \bR ^ d$ is replaced by its nearest neighbor item $m_\text{J} \in \bR ^ d$ in the memory as follows:
\begin{equation}
\label{equ_memory_replace}
\tilde{z}_i \leftarrow m_\text{J},  \text{ where } \text{J} = \text{argmin}_j \| z_i - m_j \|_2, \quad j \in \{1, \cdots, k\}.
\end{equation}
Then the input to the decoder $\bde_p$ is the substituted feature $\tilde{z}$.
To simplify the description, we denote $\cg$ as the function of retrieving the corresponding item in the memory and using the retrieved item to replace the input. Then we arrive at the following mapping function:
\begin{equation}
\tilde{z} = \cg(z)
\end{equation}

In the training phase, we update the memory with exponential moving averages \cite{van2017neural}. We denote ${z_{i,1}, z_{i,1}, \dots, z_{i, n_i}}$ as $n_i$ latent features that are closest to the memory item $m_i$. In order to make the memory item close to the set of latent features, we have the loss as:
\begin{equation}
\sum_j^{n_i}{\left\lVert z_{i, j} - m_{i} \right\rVert}^2.
\end{equation}
The optimal $m_i$ has a closed form solution, which is the average of the latent features in the set:
\begin{equation}
m_i = \frac{1}{n_i}\sum_j^{n_i}z_{i,j}.
\end{equation}
In order to handle online mini-batch training, the exponential moving average \cite{van2017neural} is used:
\begin{equation}
\begin{aligned}
N_i^{(t)} &= N_i^{(t-1)} \gamma + n_i^{(t)}(1 - \gamma) \\
e_i^{(t)} &= e_i^{(t-1)} \gamma + \sum_j z_{i,j}^{(t)}(1 - \gamma) \\
m_i^{(t)} &= \frac{e_i^{(t)}}{N_i^{(t)}},
\end{aligned}
\end{equation}
where $\gamma$ is a constant between 0 and 1, meaning how much history data to be kept. $n_i^{(t)}$ and $z_{i,j}^{(t)}$ denote $n_i$ and $z_{i,j}$ in the $t-th$ mini-batch, respectively.

During the training of Proxy Extraction Module, we update the encoder and decoder simultaneously with the update of the memory. The encoder and decoder are optimized as:
\begin{equation}
\cL_\text{p} = \| \bde_p(\cg(\ben_p(\bI))) - \bP \|_2^2,
\end{equation}
where $\bP$ is the SI extracted from the original image $\bI$ with the SLIC algorithm \cite{achanta2012slic}. It is worth noting that there is no gradient defined for the function $\cg$. However, it is possible to approximate the gradient with the straight-through estimator \cite{van2017neural}. We copy the gradient of $\tilde{z}$ as the gradient of $z$.
Once the Proxy Extraction Module is trained, the weights of $\bde_p$, $\ben_p$ and the items in memory are fixed when optimizing the Image Reconstruction Module.

\highre{
\textbf{Our method vs. MemAE \cite{gong2019memorizing}}. Recently, Gong \etal \cite{gong2019memorizing} proposed to augment an Auto-Encoder with the memory module, termed MemAE.
Given an input, the encoder in MemAE extracts an encoded representation, which is used as a query to retrieve the most relevant items in the memory. The multiple items are then averaged with an attention weight to get the substituted feature $\tilde{z}$ for the reconstruction. 
Both the MemAE and our method can take advantage of memory-augmented networks for anomaly detection.
However, the specific task in this work is relatively simpler than \cite{gong2019memorizing}. Thus, our method only retrieve the nearest item as the substituted feature rather that retrieve multiple items.
The reasons why the task in this paper is relatively simpler are in two aspects: 
first, the patterns in the medical images are simpler than the patterns in the natural images \cite{shen2020domain, shen2020modeling, pham2017brain}; second, MemAE \cite{gong2019memorizing} reconstructs original image while our method predicts the SI, which contains the simple pattern than that in original image.
As a result, if a combination of multiple memory items is used, some anomalies may still have the chance to be well reconstructed. Therefore, we only retrieve a single item rather multiple items for SI prediction.
}

\subsection{Image Reconstruction Module}
In anomaly detection, the abnormal images are usually not available in the training phase.
However, compared with the normal SI, the input image could be regarded as a pseudo abnormal SI in appearance. This inspires us to synthesize a pseudo abnormal proxy by cutting a patch from the normal image, and paste it into the normal proxy.
Then we obtain the pseudo abnormal proxy $\bPb$, and $\bM \in \{0, 1\}$ denotes a binary mask indicating the location of pasting the patch into $\bP'$. \highre{The pseudo abnormal proxy is inputted into the Image Reconstruction Module and we propose a repairing loss to enforce the output of abnormal proxy as a normal image. Thus, in the test phase, the reconstruction error corresponding to the abnormal images would be boosted, and consequently, the discrepancy between the normal images and the abnormal images is enlarged, which is a desirable feature for anomaly detection.}

\begin{figure}[htb]
	\centering
	\includegraphics[width=0.49\textwidth]{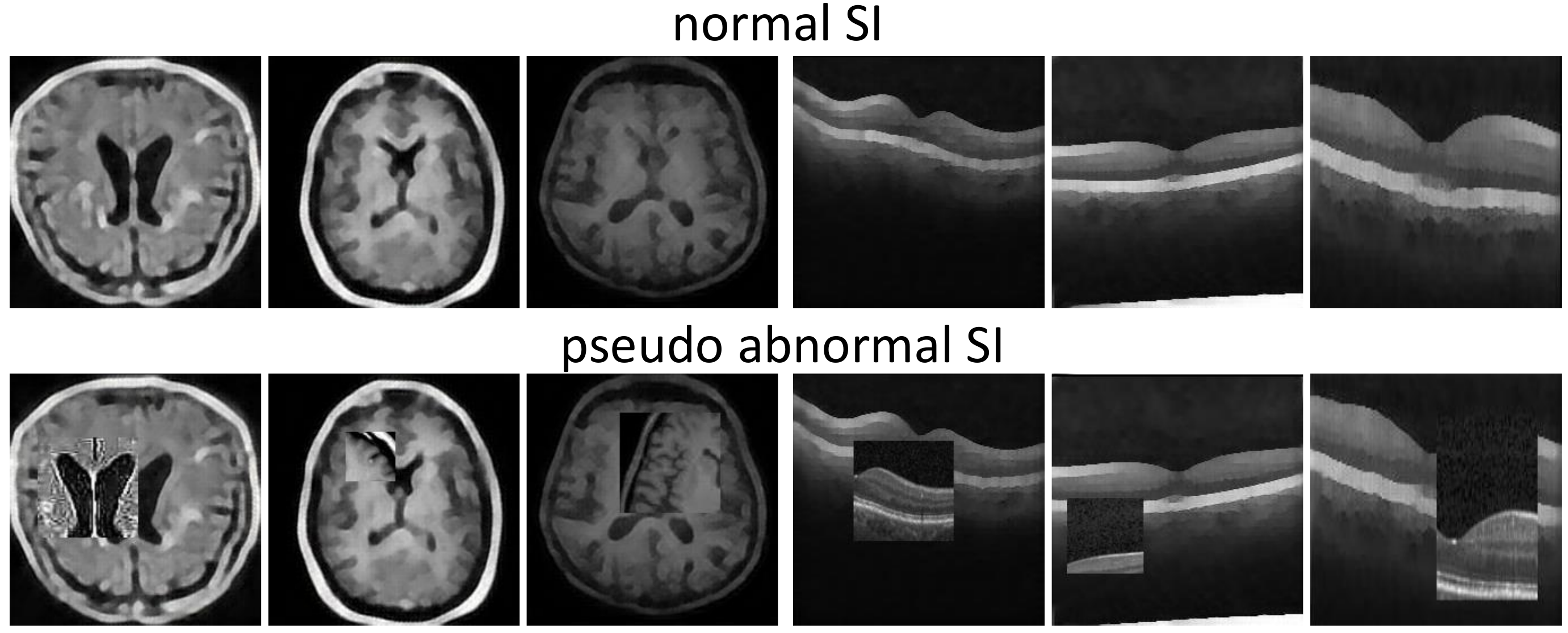}
	\vspace{-0.2in}
	\caption{
		Examples of the normal SI (first row) and the pseudo abnormal SI (second row) in brain MRI and retinal OCT. }
	\label{fig_pab}
%	\vspace{-0.10in}	
\end{figure}

To obtain diverse pseudo abnormal proxy  $\bPb$, we propose a Pseudo Abnormal Proxy Constructor (PAPC) that crops the patch from the image with a random size and pasts it into the proxy at a random position. As shown in Fig. \ref{fig_pab}, the pseudo abnormal proxy can mimic the abnormal images to some extent.
In the training phase, we enforce the network to reconstruct the normal image even with the pseudo abnormal SI, which is like the `anomalies repairing'. In this way, the reconstruction error corresponding to the abnormal images would be boosted, and consequently, the discrepancy between the normal images and the abnormal images is enlarged, which is a desirable feature for anomaly detection. 

Specifically, in the training of the Image Reconstruction Module, the forward pass of the normal proxy and the pseudo abnormal proxy is denoted as $\bhI = \cf_g(\hat{\bP})$ and $\bhIb = \cf_g(\bPb)$, respectively. The function $\cf_g(\cdot)$ is shared.

\textbf{Normal Image Reconstruction Loss.}
For the normal proxy, the image reconstruction loss is shown as follow:
\begin{equation}
\cL_{\text{rec}} = \| \bhI - \bI \|_2^2 + \lambda_{\text{g}} \underbrace{\big(\bE[\log (1-\bD(\bhI)))] + \bE[\log \bD(\bI)]\big)}_{\text{adversarial loss}},
\end{equation}
where $\bD$ denotes the discriminator \cite{goodfellow2014generative}, and the $\lambda_{\text{g}}=0.01$ is a hyper-parameter.

\textbf{Abnormal Image Repairing Loss.}
For the pseudo abnormal proxy, we propose a repairing loss to enforce the output of abnormal proxy as a normal image. The repairing loss consists of a global regularization item and a local regularization item, which are defined as follows:
\begin{equation}
\cL_{\text{global}} = \| \bhIb - \bI \|_2^2 + \lambda_{\text{g}}   \big(\bE[\log (1-\bD(\bhIb)))] + \bE[\log \bD(\bI)]\big),
\vspace{-0.20in}
\end{equation}

\begin{equation}
\begin{aligned}
\cL_{\text{local}} = & \| \bM \odot \bhIb - \bM \odot \bI \|_2^2 +  \\
& \lambda_{\text{g}} \big( \bE[\log (1-\bD(\bM \odot \bhIb)))] + \bE[\log \bD(\bM \odot \bI)] \big),
\end{aligned}
\end{equation}
where $\odot$ is element-wise multiplication.

\textbf{Total Objective Function.}
We arrive at the objective function for the Image Reconstruction Module:
\begin{equation}
\begin{aligned}
\cL = & \cL_{\text{rec}} + \cL_{\text{repairing}} \\
= & \cL_{\text{rec}} + \lambda_\text{global}   \cL_{\text{global}} + \lambda_{\text{local}}  \cL_{\text{local}},
\end{aligned}
\end{equation}
where $\lambda_\text{gobal}$ and $\lambda_\text{local}$ are the hyper-parameters. Empirically, we set $\lambda_\text{gobal}=0.25$, $\lambda_\text{local}=0.5$ on all datasets in our experiments.

\subsection{Anomaly Detection on Test Data}
Taking an original image $\bI \in \bR ^{w \times h}$ as an input, the proposed method can obtain an extracted proxy $\hat{\bP} \in \bR ^{w \times h}$ and a reconstructed image $\bhI \in \bR ^{w \times h}$. The $w$ and $h$ denote the width and the height of image, respectively. We compute the anomaly score for the image-level anomaly detection in the latent feature space, and compute the anomaly score map for the pixel-level anomaly detection in the image space.

\textbf{Anomaly Score for Image-level Anomaly Detection.}
It has been proven that computing the anomaly score by mapping the image space to the latent space is effective \cite{akcay2018ganomaly, zhou2020sparse}.
However, training an additional encoder is inefficient and redundant. We apply the existing encoder $\ben_p$  in Proxy Extraction Module to map the image to a latent space, and we get the latent feature as:
\begin{equation}
\begin{aligned}
z &= \ben_p(\bI) \\
\hat{z} &= \ben_p(\bhI)
\end{aligned},
\end{equation}
where $\hat{z}$ is the latent feature of the reconstructed image $\bhI$.

We compute the anomaly score as:
\begin{equation}
\label{equ_score}
\bA_\text{img} =  \| z - \hat{z} \|_F.
\end{equation}

\textbf{Anomaly Score Map for Pixel-level Anomaly Detection.}
To get the pixel-level anomaly map $\bA_\text{pix} \in \bR ^{w \times h} $ for lesion segmentation, we compute the anomaly score map in the image space as:
\begin{equation}
\bA_\text{pix} = | \bI - \bhI |.
\end{equation}

\subsection{Detailed Network Architecture}
\begin{figure}[htb]
	\centering
	\includegraphics[width=0.49\textwidth]{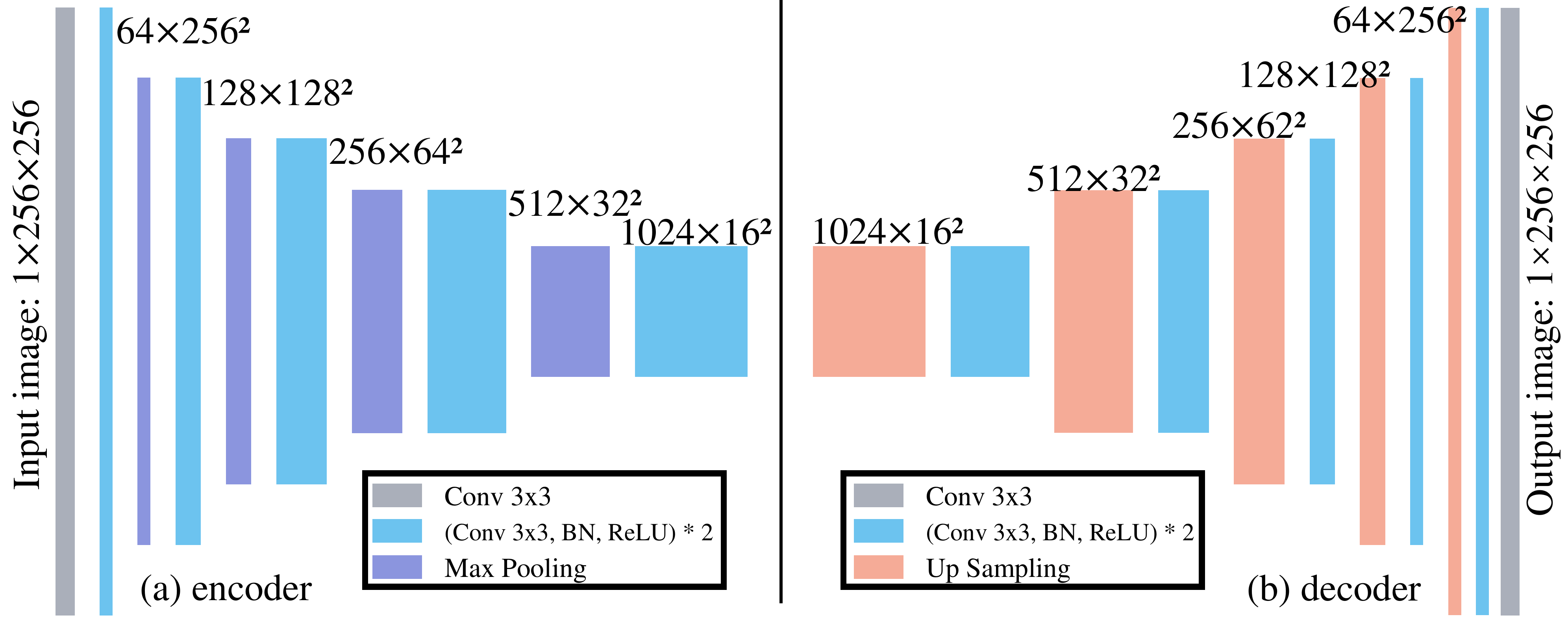}
	\vspace{-0.25in}
	\caption{The architectures of the encoder and decoder.
	}
	\label{fig_encoder_decoder}
	\vspace{-0.1in}
\end{figure}

We use the same encoder and decoder in both the Proxy Extraction Module and the Image Reconstruction Module.
The detailed architectures of the encoder and the decoder are shown in Fig. \ref{fig_encoder_decoder}.

%% file: sections/exps.tex
\section{Experiments}
\label{sec_exp}
\subsection{Experimental Setup}

\begin{table*}
	\centering
%	\normalsize
	\small
	\caption{The \textup{image-level} anomaly detection results on brain MRI and retinal OCT, and the \textup{pixel-level} anomaly detection results on retinal fundus. The best two results are shown in \first{red} and \second{red} fonts.}
	%	\vspace{-0.05in}
	\begin{tabular}{c|ccc|ccc|ccc}
		\hline
		\multirow{2}{*}{Mehtods}
		& \multicolumn{3}{c|}{Brain MRI \cite{rai2019hybrid}}
		& \multicolumn{3}{c|}{Retinal OCT \cite{kermany2018identifying}}
		& \multicolumn{3}{c}{Retinal Fundus \cite{porwal2020idrid}}\\ \cline{2-10}
		& AUC & ACC & \HILIGHT{F1-score} & AUC & ACC & \HILIGHT{F1-score}   & AUC   & ACC & \HILIGHT{F1-score}  \\ \hline
		Auto-Encoder \cite{baur2018deep}   		& 0.766 & 0.714 & 0.674 & 0.783 & 0.751 & 0.669 &  0.609 & 0.589 & 0.542                    \\
		\highre{MemAE} \cite{gong2019memorizing}   & \second{0.848} & 0.789 & \first{0.722} & 0.876  & 0.838  & 0.652 & 0.647 &  0.592 & 0.567 \\
		AnoGAN \cite{schlegl2017unsupervised}   & 0.803 & 0.757 & 0.691 &0.846 & 0.789 & 0.637 &  0.630 & 0.618  & 0.579               \\
		f-AnoGAN \cite{schlegl2019f}   			&  0.822 & 0.764 & 0.675 &0.882 & 0.808 & 0.653 & \second{0.698} & \first{0.686}  & 0.637                  \\
		GANomaly \cite{akcay2018ganomaly}   	& 0.832 & \second{0.798} & 0.667 &0.916 & 0.826 & \first{0.727} &  0.652 & 0.633& \first{0.658}                  \\
		Sparse-GAN \cite{zhou2020sparse}			& 0.835 & 0.791 & 0.645 &\second{0.925} & \second{0.841} & 0.714  & 0.663 & 0.638 & \second{0.651}  \\
		pix2pix \cite{isola2017image}   		&  0.796 & 0.737 & 0.617 &0.861 & 0.818 &  0.702 & 0.632 & 0.621 & 0.603         \\
		Cycle-GAN \cite{zhu2017unpaired}   		&  0.808 & 0.752 & \second{0.712} &0.815 & 0.762  & 0.675 & 0.626 & 0.613 & 0.608         \\ \hline
		\textbf{Our Method} &  \first{0.853} & \first{0.805} & 0.709 &\first{0.933} & \first{0.849}  & \second{0.725} & \first{0.701} & \second{0.682}& 0.649  \\ \hline
	\end{tabular}
	\label{table1}
	%	\vspace{-0.05in}
\end{table*}

\subsubsection{Implementation Details}
The optimizer we used is the Adam optimizer \cite{kingma2014adam}, and the learning rate is set as 0.001. 
The scale of original image and its proxy are resized to be 256 $\times$ 256. The number of superpixels is set to be 800 in the SLIC \cite{achanta2012slic} algorithm.
In the memory, we set $d$ = 64, $k$ = 128, $\gamma$ = 0.99. 
In our implementation, the Proxy Extraction Module and the Image Reconstruction Module are trained separately, and we find that training two modules within an end-to-end learning manner does not necessarily bring performance gain. 

\subsubsection{Evaluation Metrics}
For performance evaluation, we calculate the Area Under Receiver Operation Characteristic (AUC) by gradually changing the threshold of anomaly score $\bA_\text{pix}$ and $\bA_\text{img}$ 
%for normal/abnormal detection 
for the pixel-level and the image-level anomaly detection, respectively. By following \cite{zhou2020sparse, wolleb2020descargan}, we also calculate the accuracy (ACC) and \HILIGHT{F1-score} for the performance evaluation. \HILIGHT{The F1-score is defined as the harmonic mean of precision and recall of a model.
}

\subsubsection{Baseline Methods}
We compare our method with several state-of-the-art anomaly detection methods, including 
\highre{MemAE} \footnote{\url{https://github.com/donggong1/memae-anomaly-detection}} \cite{gong2019memorizing},
AnoGAN \footnote{\url{https://github.com/LeeDoYup/AnoGAN-tf}} \cite{schlegl2017unsupervised}, f-AnoGAN \footnote{\url{https://github.com/tSchlegl/f-AnoGAN}} \cite{schlegl2019f}, GANomaly \footnote{\url{https://github.com/samet-akcay/ganomaly}} \cite{akcay2018ganomaly}, Sparse-GAN \cite{zhou2020sparse}, Auto-Encoder \cite{baur2018deep} based anomaly detection method, and pix2pix \footnote{\url{https://github.com/junyanz/pytorch-CycleGAN-and-pix2pix}} \cite{isola2017image} based anomaly detection method. 
As our method consists of the translation between image and proxy, we also compare our ProxyAno with Cycle-GAN $^\text{\textcolor{red}{5}}$ \cite{zhu2017unpaired}.
\highre{We use the code provided in the GitHub to implement the baseline methods. For a fair comparison, the architecture of the encoder and decoder in Auto-Encoder is the same as that used in our method.}

\subsection{Image-level Anomaly Detection on Brain MRI}
\subsubsection{\textbf{Dataset}}
The training dataset used in previous brain MRI anomaly detection work \cite{baur2018deep, chen2018unsupervised} are not released. However, there are several publicly available brain MRI datasets \cite{rai2019hybrid, sartaj2020brain} used for tumor classification and lesion segmentation.
Thus, we propose to integrate \cite{rai2019hybrid} and \cite{sartaj2020brain} and use it to evaluate our proposed method.
There are 500 normal images in the training set, and the test set consists of 500 abnormal image and 500 normal images.

\begin{figure*}[ttt]
	\centering
	\includegraphics[width=.89\textwidth]{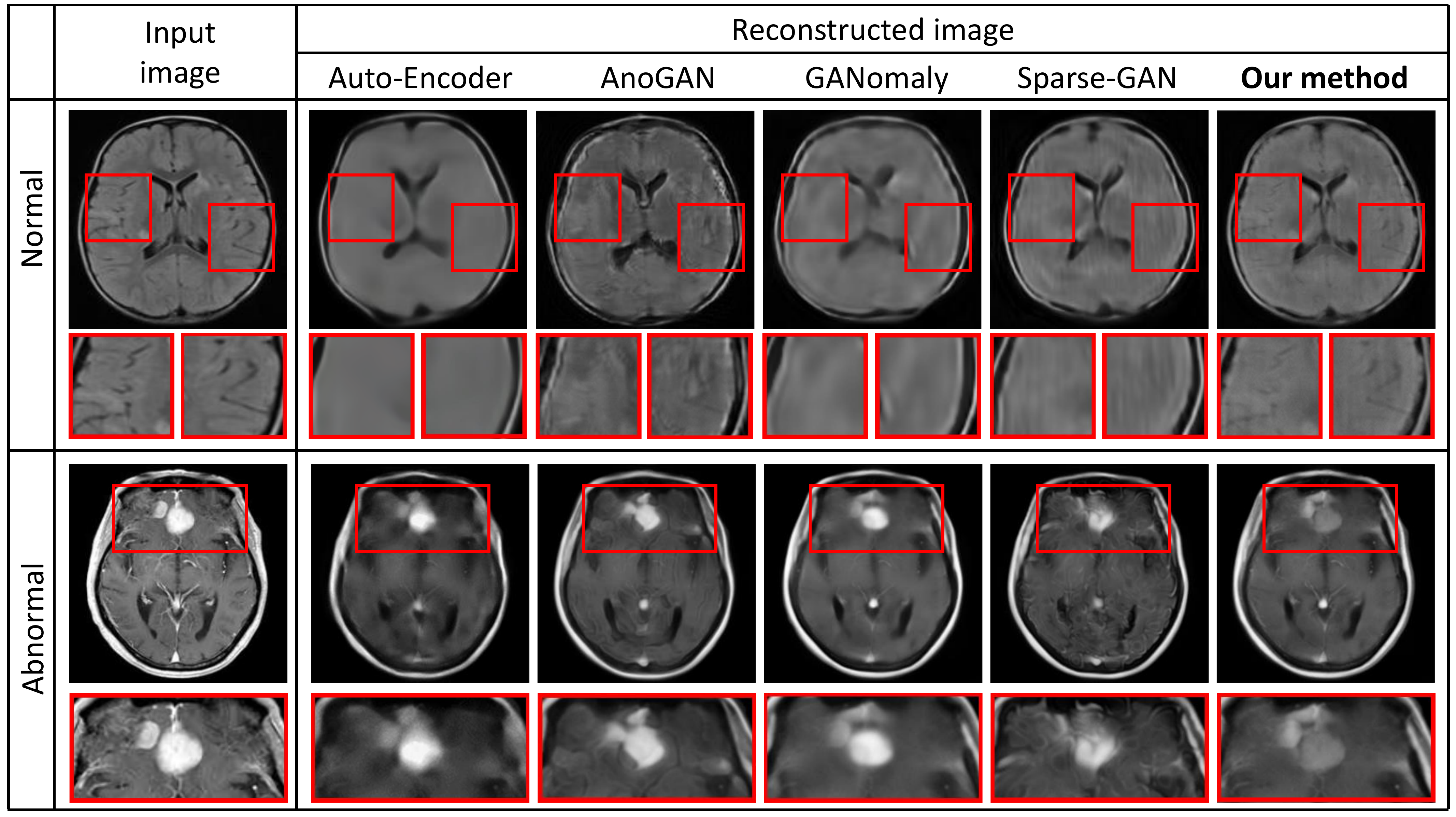}
	\vspace{-0.12in}
	\caption{The reconstruction results on the normal and the abnormal brain MRI. The red rectangle in the abnormal sample denotes the brain tumor. From left to right: input brain MRI, reconstruction result obtained by Auto-Encoder \cite{baur2018deep}, AnoGAN \cite{schlegl2017unsupervised}, GANomaly \cite{akcay2018ganomaly}, Sparse-GAN \cite{zhou2020sparse} and our method. Our method reconstructs the abnormal regions poorly while reconstructing the normal image well.
	}
	\label{fig_baselines}
	\vspace{-0.15in}
\end{figure*}

\subsubsection{\textbf{Performance Evaluation}}
We report the performance of different methods for image-wise anomaly detection on the brain MRI dataset in Table \ref{table1}.
We can see that the AUC and ACC of our proposed ProxyAno outperforms that of all baseline methods on the brain MRI dataset. In particular, ProxyAno achieves an AUC of 0.853 on the brain MRI dataset with an increase of 1.8\% from 0.835 by the latest Sparse-GAN \cite{zhou2020sparse}. \HILIGHT{However, the F1-score of our method does not achieve the best. From Table \ref{table1}, the similar results can be found in other two datasets \cite{kermany2018identifying}\cite{porwal2020idrid}, \ie, the AUC of our method achieves the best while the F1-score of our method does not. The reason behind this is probably the threshold selection has an important effect on the F1-score while it dose not effect the AUC result.}

We also show the reconstruction results in Fig. \ref{fig_baselines} to visually compare our method with other competitive methods, including Auto-Encoder \cite{baur2018deep}, AnoGAN \cite{schlegl2017unsupervised}, GANomaly \cite{akcay2018ganomaly} and Sparse-GAN \cite{zhou2020sparse}. The images show that our method can well reconstruct the normal images while poorly reconstruct the abnormal ones.

\subsection{Image-level Anomaly Detection on Retinal OCT}

\subsubsection{\textbf{Dataset}}
The retinal OCT \cite{kermany2018identifying} dataset from Spectralis OCT (Heidelberg Engineering, German) is used in this section, and it contains data with three different lesions: drusen, DME (diabetic macular edema), and CNV (choroidal neovascularization).
This dataset contains the standard training/test split. We use the normal images in the original training set to train the model, and use all the test images for performance evaluation.

\subsubsection{\textbf{Performance Evaluation}}
As shown in the Table \ref{table1}, our method outperforms all baseline methods on the retinal OCT dataset.
Particularly, the proposed ProxyAno achieves an AUC of 0.933 in the retinal OCT. 
Further, our method significantly outperforms the Auto-Encoder based solution because our solution can mitigate the identity mapping issue and increase the reconstruction error for the abnormal images. Our method also outperforms the Cycle-GAN based solution because we introduce the pseudo abnormal SI's, which would be repaired as the normal images. Consequently, the reconstruction errors for the abnormal images would be large, which would also facilitate the anomaly detection.

We further compute the average anomaly score for the image-level anomaly detection for both the normal images and the abnormal images on the OCT dataset with Equation (\ref{equ_score}).
Then, we calculate the gap of these two scores to measure the ability of our method and other baselines to discriminate the normal and the abnormal images. A larger gap means the normal and the abnormal images can be more easily separated.
The results in Table \ref{table_gap} show that our method achieves the largest gap than other baselines, which validates the effectiveness of our method for anomaly detection.

\begin{table}[hhh]
	\centering
	\caption{The average anomaly score for the normal images, the abnormal images and the gap between these two scores on the retinal OCT dataset. \highre{Our SI approach denotes using SI to connect two primitive encoder-decoder pairs, without using memory and repairing loss.}}
	\vspace{0.03in}
	\begin{tabular}{c|c|c|c}
		\hline
		\multirow{2}{*}{\qquad\quad Method \qquad\quad} & \multicolumn{3}{c}{Anomaly Score} \\ \cline{2-4} 
		& normal     & abnormal     & \quad\, gap \quad\,    \\ \hline
		Auto-Encoder \cite{baur2018deep} 	& 0.601 & 0.898	& 0.297  \\
		f-AnoGAN \cite{schlegl2019f} 		& 0.549 & 0.922	& 0.373     \\ 
		GANomaly \cite{akcay2018ganomaly} 	& 0.430 & 0.834 & 0.404		\\
		Sparse-GAN \cite{zhou2020sparse} 	& 0.576 & 0.995 & 0.419 	\\ \hline
		\highre{Our SI approach}			& 0.580 & 1.006 & 0.426 \\
		Our Final Model 					& 0.568 & 1.003 & \textbf{0.435} \\
		\hline
	\end{tabular}
	\label{table_gap}
\end{table}

%\begin{table}[hhh]
%	\centering
%	\caption{The average anomaly score for the normal images, the abnormal images and the gap between these two scores on the retinal OCT dataset.}
%	\begin{tabular}{c|c|c|c}
%		\hline
%		\multirow{2}{*}{\qquad\quad Method \qquad\quad} & \multicolumn{3}{c}{Anomaly Score} \\ \cline{2-4} 
%		& normal     & abnormal     & \quad\, gap \quad\,    \\ \hline
%		f-AnoGAN \cite{schlegl2019f} 		& 0.549 $\pm$ 0.016 & 0.922	$\pm$ 0.017	& 0.373     \\ 
%		GANomaly \cite{akcay2018ganomaly} 	& 0.430 $\pm$ 0.022 & 0.834 $\pm$ 0.021 & 0.404		\\
%		Sparse-GAN \cite{zhou2020sparse} 	& 0.576 $\pm$ 0.023 & 0.995 $\pm$ 0.010 & 0.419 	\\ \hline
%		Our Method 							& 0.568 $\pm$ 0.018 & 1.003 $\pm$ 0.013 & \textbf{0.435} \\
%		\hline
%	\end{tabular}
%	\label{table_gap}
%\end{table}

\begin{figure*}[htb]
	\centering
	\includegraphics[width=0.9\textwidth]{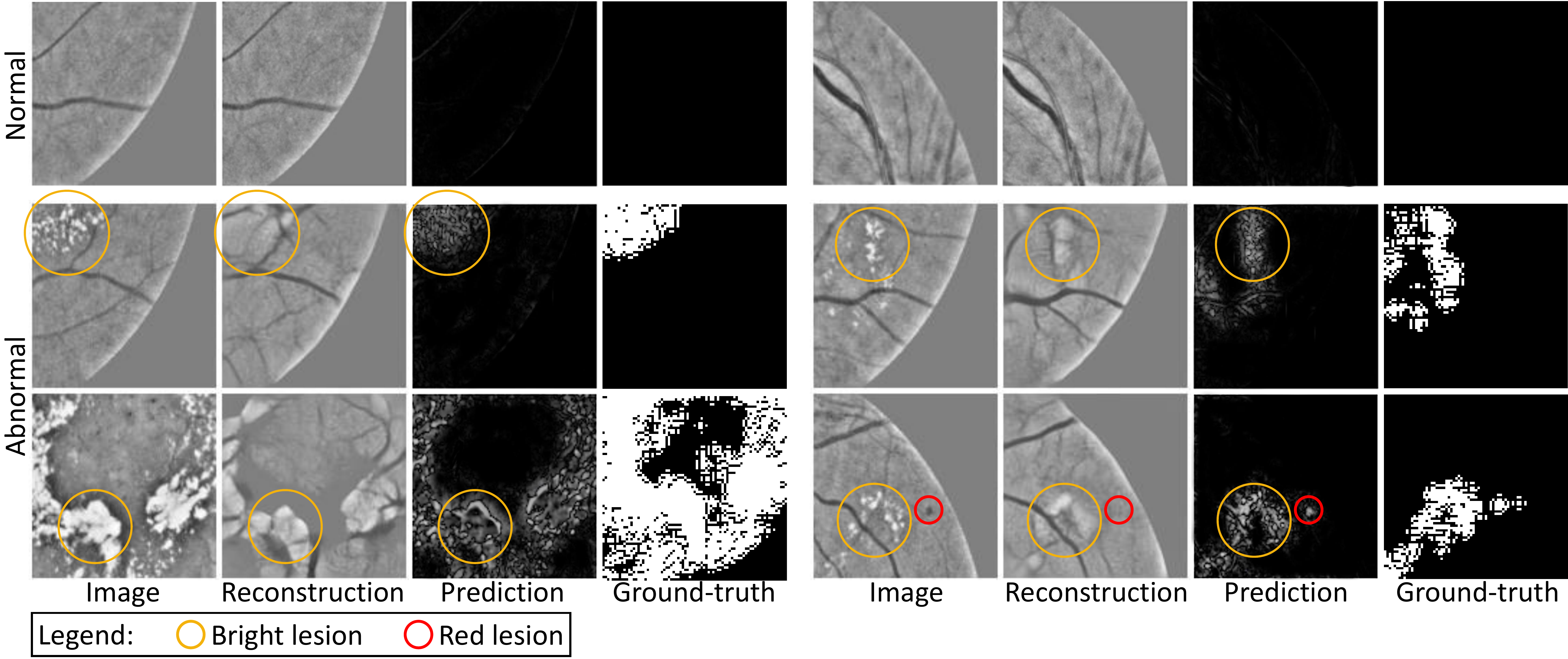}
	\vspace{-0.10in}
	\caption{The qualitative results on the retinal fundus image. Both the red lesions and bright lesions can be roughly detected with our method.
	}
	\label{fig6}
	\vspace{-0.15in}
\end{figure*}

\subsection{Pixel-level Anomaly Detection on Retinal Fundus Image}
\label{sec_idrid}

\subsubsection{\textbf{Dataset}} Indian Diabetic Retinopathy Image Dataset (IDRiD) \cite{porwal2020idrid} is a publically available dataset used to evaluate Diabetic Retinopathy (DR) detection. To train the proposed model, we only choose the normal class from the original training set in IDRiD as our training set, and the test set is the same as the original lesion detection dataset.
IDRiD consists of 134 normal images for training, and 69 abnormal images with pixel-level lesion annotation for testing.
Since the resolution of original image is very large (i.e., $4288 \times 2848$), we crop 9 (i.e., $3 \times 3$) patches from each image.

\subsubsection{\textbf{Performance Evaluation}}
The baseline methods for pixel-level anomaly detection are the same as that in image-level performance evaluation. To compute the lesion detection accuracy, we set the threshold of $\bA_\text{pix}$ as 0.5. As shown in Table \ref{table1}, the AUC and ACC of proposed method outperforms the baseline methods, which further verifies the effectiveness of our method.
We further qualitatively show the reconstruction and lesion detection results in Fig. \ref{fig6}.
It can be found that the normal region is reconstructed with a small error, while the region with lesion is reconstructed with a large error. Both the red lesions and bright lesions can be roughly detected.
Here the red lesions include haemorrhages of all shapes and microaneurysms, bright lesions include hard and soft exudates, drusen, cotton-wool spots \cite{playout2019novel}.

\subsection{The Selection of The Proxy}
\label{sec_super_works}

\begin{figure}[htb]
	\centering
	\includegraphics[width=.47\textwidth]{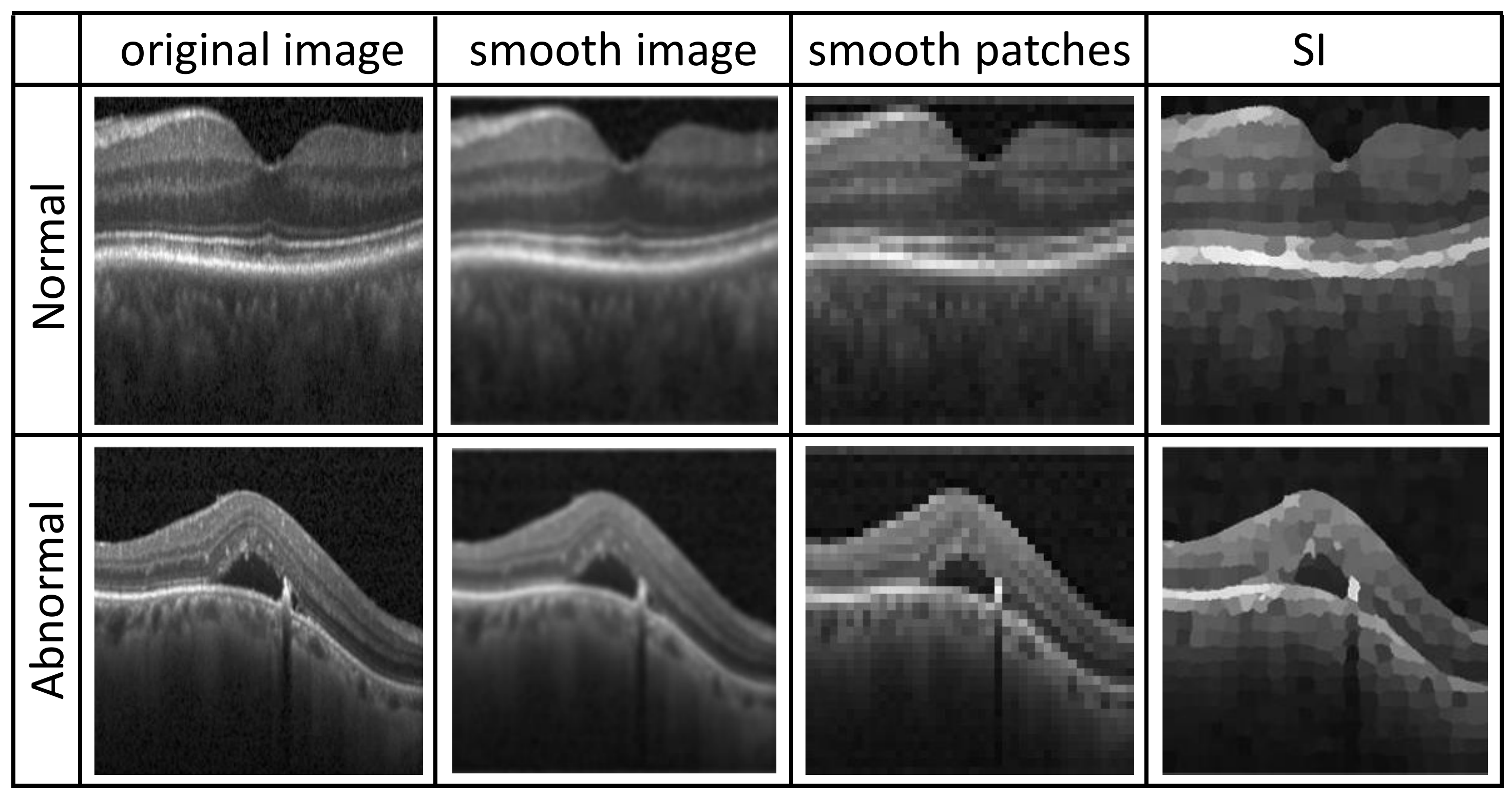}
	\vspace{-0.1in}
	\caption{\highre{Illustration of different types of proxy. Compared with smooth image and smooth patches (short for image with smooth patches), the SI preserves local structure.}}
	\label{fig_proxies}
	\vspace{-0.05in}	
\end{figure}

\highre{
	In this part, to verify the effectiveness of proxy approach, we study different proxy types and compare them with baseline that without proxy (\ie, Auto-Encoder).
	Specifically, besides the SI, we also consider the following five types of proxy:
	1) the edge extracted by a Canny edge detector;
	2) smooth image, which is a short for the image smoothed with Gaussian blur;
	3) image with smooth patches, which denotes the image consists of patches colored with the average intensity in each patches;
	4) edge $\oplus$ smooth image, here $\oplus$ denotes the concatenation;
	5) edge $\oplus$ image with smooth patches.
	The experiments are conducted on the retinal OCT \cite{kermany2018identifying} dataset.
	The illustration of smooth image and image with smooth patches can be found in Fig. \ref{fig_proxies}.
}

\begin{figure*}[htb]
	\centering
	\includegraphics[width=.9\textwidth, height=4.3in]{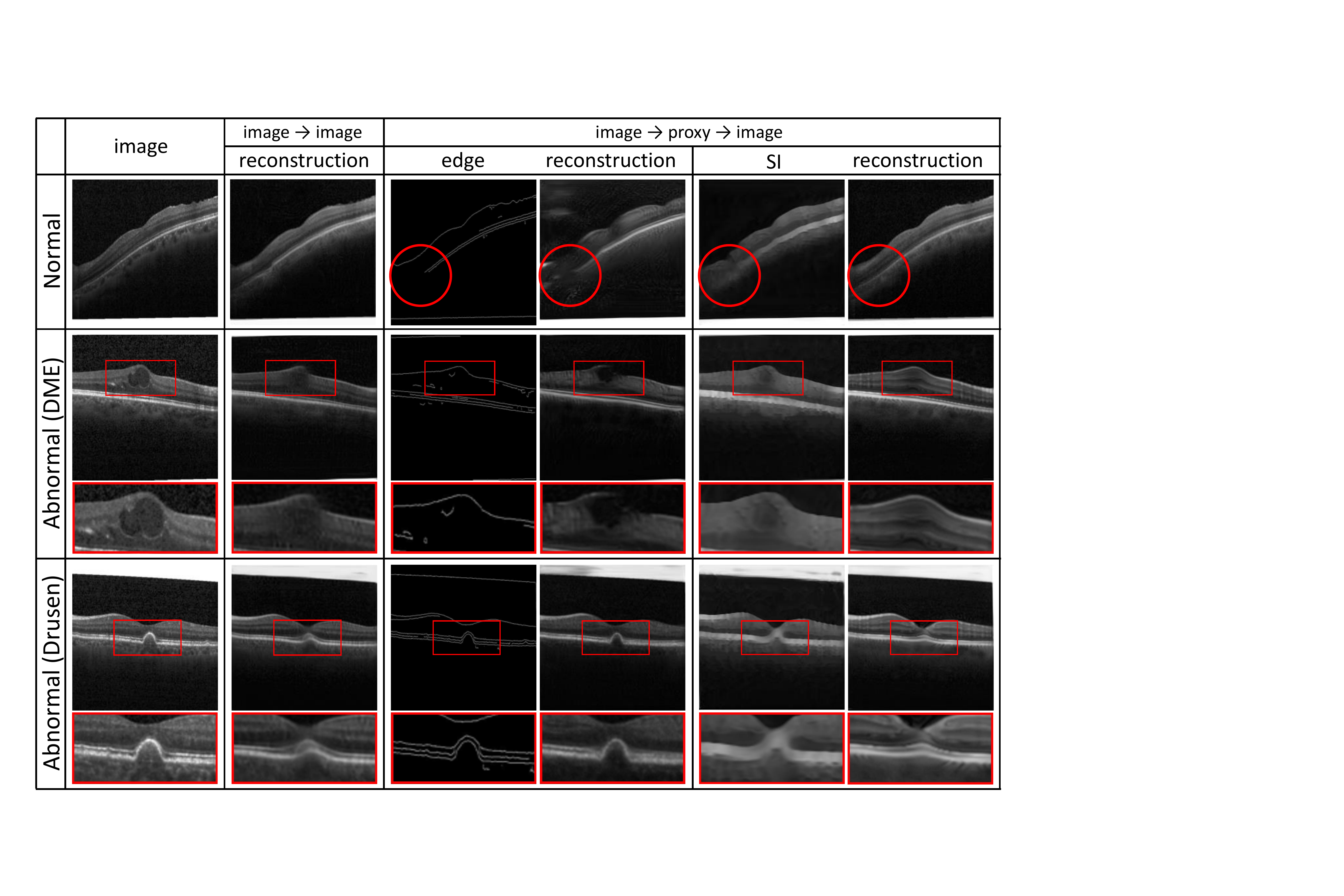}
	\vspace{-0.05in}
	\caption{The visualization results with different proxy.}
	\vspace{-0.08in}
	\label{fig_str_spi}
\end{figure*}

%\begin{figure*}[htb]
%	\centering
%	\includegraphics[width=.85\textwidth]{figures/fig_texture}
%	\caption{\highre{The qualitative results of different texture categories. The texture anomaly in wood and grid can be roughly detected, while the texture anomaly in carpet and tile cannot be detected, which show the limitation of the proposed method.}}
%	\label{fig_texture}	
%\end{figure*}

The quantitative results are reported in Table \ref{table_str_spi}, from which we can see that all proxy methods outperform the baseline without proxy, validating proxy approach is effective. We can also see that the SI-based proxy is superior to all other types of proxy.
The reason of SI-based proxy outperforms the edge-based proxy is possibly because that the edge contains less texture information, and makes the image reconstruction difficult. 
%A similar argument can also be found in previous work \cite{han2019finet, ren2019structureflow, zhao2020uctgan}, which show the multiple possibilities of the images reconstructed merely based on edges (or sketches). Reconstruction with multiple possibilities means that it is hard to accurately reconstruction. 
Compared with the edges, the proposed SI contains more texture information (\ie, the average intensity within each superpixel), which makes the image reconstruction easier, and facilitates the anomaly detection. 
\highre{
	The reason of SI-based proxy outperforms the proxy based on smooth image and image with smooth patches is possibly because that the SI contains more structure information. 
	Compare with the smooth image, in the SI the edge is relatively enhanced. Compare with the image with smooth patches, the SI contains more semantic information and preserves local structure. 
	As a summary, compare with other types of proxy, the proposed SI contains more texture and structure information, validating that both texture and structure information are important for anomaly detection. This similar argument can also be found in 
	\cite{zhou2020encoding}. Moreover, with concatenating edge, the performance of both smooth image and image with smooth patches improved. This also validate the argument that both texture and structure information are important for anomaly detection.
}

\begin{table}[hhh]
	\centering
	\caption{\highre{The results of different proxy selection strategies.}}
	\vspace{-0.05in}
	\begin{tabular}{c|c|c}
		\hline
		\; Index \qquad & \qquad\quad\ Different Proxies \qquad\qquad &  AUC   \\ \hline
		0 & without proxy (Auto-Encoder \cite{baur2018deep}) & 0.783      \\ \hline
		1 & edge & 0.804     \\
		2 & smooth image & 0.801     \\
		3 & image with smooth patches & 0.796     \\
		4 & edge $\oplus$ smooth image & 0.813      \\
		5 & edge $\oplus$ image with smooth patches & 0.815     \\ \hline
		6 & \textbf{proposed SI}  & \textbf{0.818}         \\
		\hline
	\end{tabular}
	\label{table_str_spi}
\end{table}

The qualitative results are shown in Fig. \ref{fig_str_spi}. As shown in the images in the $1^{st}$ row of Fig. \ref{fig_str_spi}, compared with the edge-based reconstruction, our SI-based solution reconstruct the normal image more accurately.
Furthermore, the images in the $2^{nd}$ column of Fig. \ref{fig_str_spi} show that although the self-reconstruction based method (\ie, `image $\rightarrow$ image' using the AE) is trained well on the normal OCT images, when we feed the abnormal images (\ie, DME image and drusen image) to the trained AE, it can also well reconstruct the lesions in the abnormal images.
On the contrast, if we use an SI as the proxy, the issue is eased.
Specifically, as shown by the images in the $5^{th}$ column, since the SI is obtained by a neural network that trained on the normal samples, the network cannot well extract the SI on the DEM images and the drusen images, especial the lesion regions in the abnormal images. Based on the poorly extracted SI, the abnormal images cannot be well reconstructed.
The images in the $6^{th}$ column show that the lesions in the DEM image and the drusen image tend to be repaired, thus the reconstruction error between input image and reconstructed image is larger than self-reconstruction based method.

\subsection{Ablation Studies}

In this part, we conduct several ablation studies on retinal OCT dataset \cite{kermany2018identifying} to evaluate each component of our method. 
\highre{
We report the results of the difference components in Table \ref{table_ablation}.
Let EncDec denotes the encoder-decoder architecture. 
The relationship between EncDen and Auto-Encoder is: if the output of EncDec is the same as the input, the EncDec is actually the Auto-Encoder; otherwise, the EncDec is not the Auto-Encoder. The mem$^*$ denotes the memory implemented in \cite{gong2019memorizing}, while the mem denotes the memory implemented in this work. The specific difference between mem$^*$ and mem is described in the Section \ref{memory}, Proxy Extraction Module.
``EncDec $+$ mem" (or mem$^*$) denotes that there is a memory between the encoder and decoder in the Auto-Encoder. 
Both Proxy Extraction Module and Image Reconstruction Module contain the EncDec, thus we use ``$2\times$EncDec" to simply denote Proxy Extraction Module and Image Reconstruction Module. Between these two modules, we use SI as the intermediate proxy, which is denoted as ``$2\times$EncDec $+$ SI". Based on ``$2\times$EncDec $+$ SI", ``$+$ mem" denotes the Proxy Extraction Module is equipped with the memory, while ``$+$ rep" denotes the Image Reconstruction Module is equipped with the proposed repairing loss. ``$+$ lat" denotes that we compute the anomaly score in the latent space rather than in the image space.
}

\begin{table}[hhh]
	\centering
	%	\scriptsize
	\small
	\caption{\highre{The ablation studies of proposed method.}}
	\begin{tabular}{c|l|c}
		\hline	
		Index & Method & AUC  \\ \hline
		1 & EncDec (Auto-Encoder \cite{baur2018deep})		& 0.783        \\
		2 & EncDec $+$ mem$^*$ (MemAE \cite{gong2019memorizing}) & 0.816 \\ 
		3 & EncDec $+$ mem  & 0.820 \\ \hline
		4 & $2\times$EncDec $+$ SI & 0.818 \\
		5 & $2\times$EncDec $+$ SI $+$ mem & 0.854 \\
		6 & $2\times$EncDec $+$ SI $+$ rep & 0.869 \\
		7 & $2\times$EncDec $+$ SI $+$ mem $+$ rep & 0.916 \\
		8 & $2\times$EncDec $+$ SI $+$ mem $+$ rep $+$ lat (final) & \textbf{0.933} \\ \hline
	\end{tabular}
	%		\vspace{-0.10in}
	\label{table_ablation}
\end{table}

\highre{
\subsubsection{\textbf{The effectiveness of memory in Auto-Encoder}}
We first validate the effectiveness of the memory between the encoder and decoder in Auto-Encoder. As shown in row 1, row 2, and row 3 in Table \ref{table_ablation}, both memory implemented by \cite{gong2019memorizing} and implemented by ours outperform the Auto-Encoder, validating that explicitly memorize the correspondence in the feature space with memory is effective. Besides, the results of the memory implemented by ours is better than that implemented by \cite{gong2019memorizing}, validating that our memory is more suitable for anomaly detection in medical image. The reason behind this is: MemAE \cite{gong2019memorizing} retrieves multiple items while our memory only retrieves the nearest item, and combining multiple items result in some anomalies may still being well reconstructed; therefore, our memory is better than MemAE \cite{gong2019memorizing} for anomaly detection in medical image.
}

\highre{
\subsubsection{\textbf{The effectiveness of using SI as the proxy}}
The results in Table \ref{table_ablation} show that using SI as the intermediate proxy (both with and without memory) consistently achieves higher AUC than Auto-Encoder.
By comparing row 4 versus row 1 and row 5 versus row 3, it can be found that using SI as the intermediate proxy can roughly improve 3.5\% AUC. This proves that it is necessary to use SI as the proxy.
Additionally, the results in row 4 and row 2 show that our SI approach outperforms MemAE \cite{gong2019memorizing}.
}

\begin{figure}[htb]
	\centering
	\includegraphics[width=.45\textwidth]{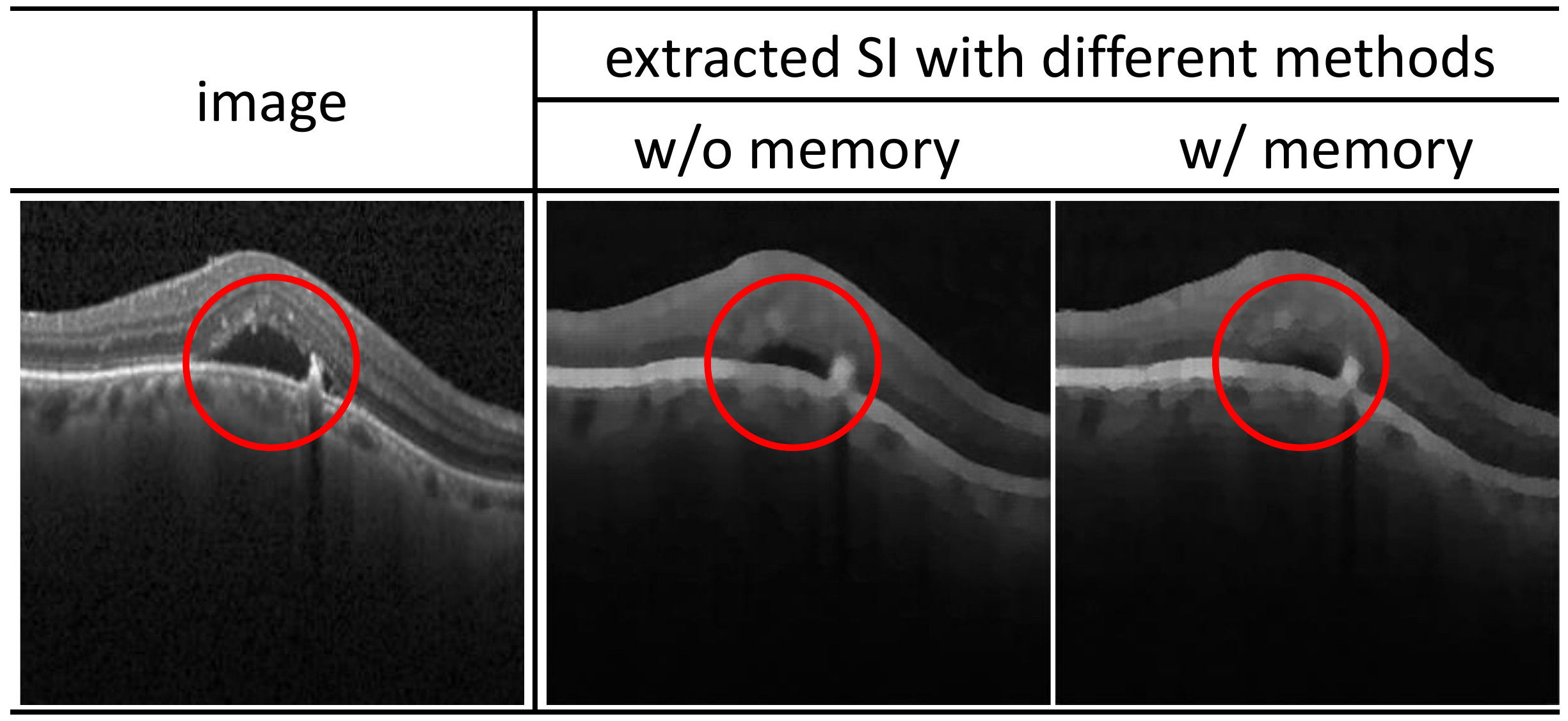}
	%	\vspace{-0.1in}
	\caption{The qualitative results of extracted SI with different methods.
		The red circle denotes the DME lesion.
		Our method tries to fill the lesion hole with normal patches. Consequently, our solution can easily identify the abnormal images.}
	\label{fig_memory_images}
	%	\vspace{-0.10in}
\end{figure}

\subsubsection{\textbf{The effectiveness of memory in Proxy Extraction Module}}
%Other than the memory-based proxy extraction, we also report the performance of an Encoder-Decoder based strategy where the output of the encoder is directly fed into the decoder for the SI extraction. 
Based on ``$2\times$EncDec $+$ SI", we equip the Proxy Extraction Module with memory to verify its effective. As shown the row 5 in Table \ref{table_ablation}, adding the memory increases the performance. 
Furthermore, we show the qualitative results of extracted SI in Fig. \ref{fig_memory_images}, and it can be observed that the DME lesion in the memory-based method is similar to the normal SI. 
This agrees with our assumption that memory makes the abnormal images look like the normal ones, which enlarges the reconstruction error of the abnormal images and facilitates the anomaly detection.

%\highre{
\subsubsection{\textbf{The effectiveness of the repairing loss in Image Reconstruction Module}}
The results in Table \ref{table_ablation} show that using the proposed repairing loss (both with and without memory) consistently achieves higher AUC than that without repairing loss.
By comparing row 6 versus row 4 and row 7 versus row 5, it can be found that trained with repairing loss can roughly improve 5\% AUC. 
This proves the effectiveness of the repairing loss in Image Reconstruction Module.
%}

\begin{table}[hhh]
	\centering
	\small
	\caption{The AUC results of different ways for anomaly score calculating.}
	\begin{tabular}{c|c|c}
		\hline
		& \quad Brain MRI \quad & \quad Retinal OCT \quad \\ \hline
		image space  & 0.807 & 0.916      \\ \hline
		latent feature space & \textbf{0.853} & \textbf{0.933}      \\ \hline
	\end{tabular}
	\label{table_feature_space}
\end{table}

\subsubsection{\textbf{The way of calculating anomaly score for image-level anomaly detection}}
\highre{In this part, we compare the way of calculating anomaly score in image space with that in latent feature space. Given an input image $\bI$ and its reconstructed counterpart $\bhI$, we can calculate the anomaly score in the \textbf{image space} based on the difference between them ($\|\bI-\bhI\|_F$).
	We compute the anomaly score in the \textbf{latent feature space} with Equation (\ref{equ_score}). }
The results are reported in Table \ref{table_feature_space}, we can see that the anomaly score calculated in the latent feature space always corresponds to the better performance. The possible reason is that the reconstructed image always contain some noises, which may affect the anomaly detection. While the latent feature can be less affected by the noises, and boosts the anomaly detection. It is worth noting that previous work \cite{akcay2018ganomaly, schlegl2019f, zhou2020sparse} also use the latent feature to calculate the anomaly score. 

\subsection{Hyper-parameters Analysis}

\begin{figure}[htb]
	\centering
	\includegraphics[width=.49\textwidth]{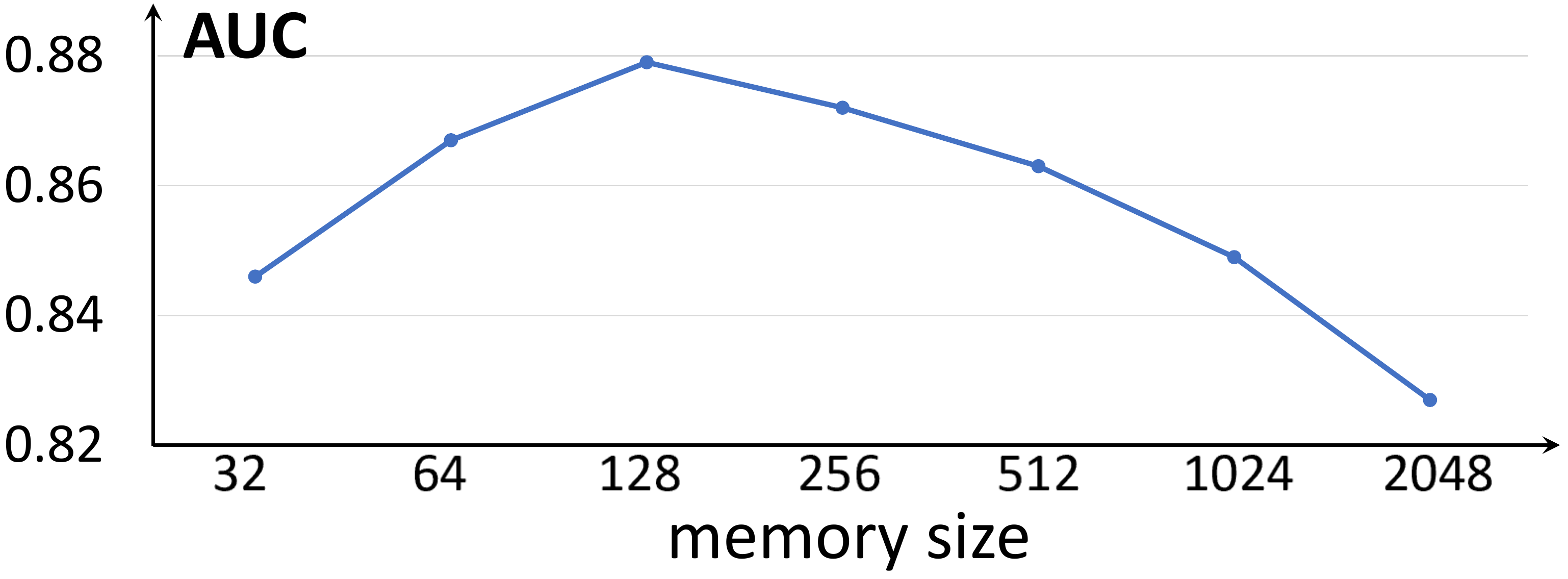}
	\vspace{-0.25in}
	\caption{The AUC results of our Proxy Extraction Module with different memory size. When setting the memory size as 2048, the model degenerates to an Encoder-Decoder model without memory.}
	\label{fig_memory_size}
	\vspace{-0.10in}	
\end{figure}

\subsubsection{\textbf{The effect of memory size}}
We gradually change the 
memory size
and show the results in Fig. \ref{fig_memory_size}, and it shows that when $k=128$ the model achieves best performance. 
The memory with a small size is incapable to characterize the mapping between the normal input and its SI, while the memory with large size cannot repair the tissues in abnormal images in the SI extraction. Theoretically, a memory with an infinite size corresponds to a model without the memory. 
Particularly, when setting the memory size as 2048, the performance of AUC decreased to 0.837, which is close to the baseline model (Encoder-Decoder w/o memory) that achieves 0.828. This also verifies the effectiveness of memory for anomaly detection.  

\begin{figure}[hhh]
	\centering
	\includegraphics[width=0.5\textwidth]{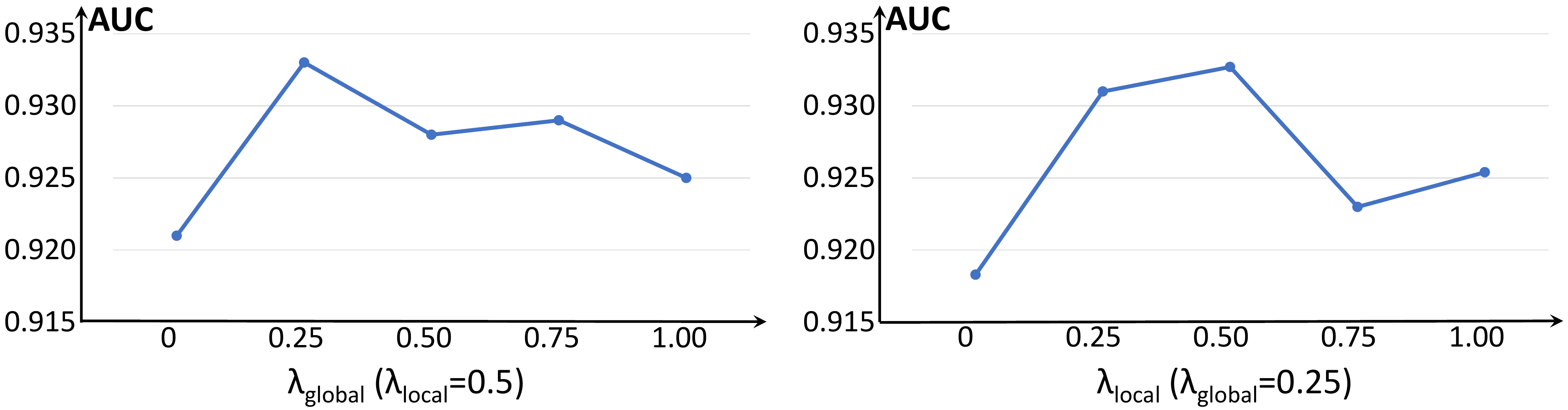}
	\vspace{-0.20in}
	\caption{The AUC results of the model with different weights. }
	%	\todo{top-bottom}
	\label{fig_repair_weight}
	\vspace{-0.10in}
\end{figure}

\subsubsection{\textbf{The pseudo anomaly repairing loss}}
\begin{figure}[hhh]
	\centering
	\includegraphics[width=0.49\textwidth]{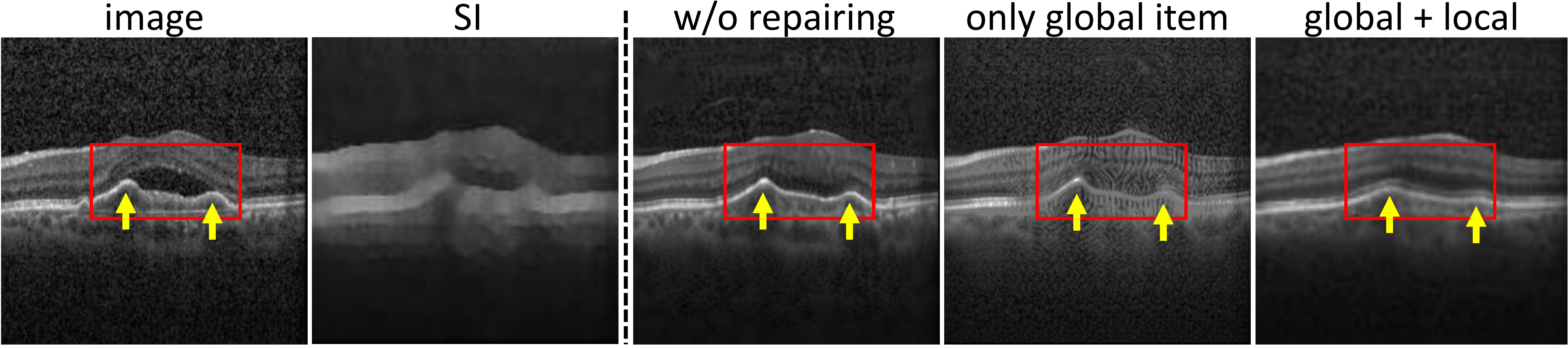}
	\vspace{-0.25in}
	\caption{The visualization results of the model with different training loss. The red rectangle denotes the DME lesion. }
	\label{fig_repair}
		\vspace{-0.10in}
\end{figure}

In our implementation, the weight for the global repairing term and local repairing term is 0.25 and 0.5, respectively. 
Then we fix the weight of one term and change the weight of another term.
As shown in the in Fig. \ref{fig_repair_weight}, we can see that the results with only local term ($\lambda_\text{global} = 0$) or global term ($\lambda_\text{local} = 0$) are worse than the results based on both terms. 
We further show the qualitative results in Fig. \ref{fig_repair}, and it can be observed that the lesion is repaired when both global and local terms are used, which verifies the effectiveness of these terms for the anomaly detection.

%\subsubsection{\textbf{Weigh in Superpixel}}
%\todo{R1-C2-Q1}

%%% old %%%
%\begin{table}[hhh]
%	\centering
%	\scriptsize
%	\caption{The ablation studies of the Proxy Extraction Module and the Image Reconstruction Module.}
%	\vspace{-0.05in}
%	\begin{tabular}{c|c|c|c}
%		\hline	
%		Index & Proxy Extraction Network & Image Generation Network & AUC  \\ \hline
%		1 & Encoder-Decoder w/o memory  & w/o repairing		& 0.818        \\
%		2 & Encoder-Decoder w/ memory   & w/o repairing		& 0.854       \\
%		3 & Encoder-Decoder w/ memory   & w/  repairing  	& \textbf{0.869}      \\  \hline
%	\end{tabular}
%		\vspace{-0.10in}
%	\label{table_spc}
%\end{table}